\documentclass[12pt,a4paper]{JHEP3}
\preprint{Cavendish-HEP-11/03\\
CERN-PH-TH/2011-054 \\
DAMTP-2011-17}

\usepackage{cite}
\usepackage{amssymb}
\usepackage{amsmath}
\usepackage{enumerate}
\usepackage{slashed}
\usepackage{graphicx}
\newcommand{\beq}{\begin{equation}}
\newcommand{\eeq}{\end{equation}}
\newcommand{\beqn}{\begin{eqnarray}}
\newcommand{\eeqn}{\end{eqnarray}}
\newcommand{\bmat}{\begin{pmatrix}}
\newcommand{\emat}{\end{pmatrix}}
\def\vp{\mathbf{p}}
\def\svp{\slashed{\vp}}
\def\vn{\mathbf{n}}
\def\vb{\mathbf{b}}
\title{Polynomials, Riemann surfaces, and reconstructing missing-energy events}

\author{Ben Gripaios$^1$, Kazuki Sakurai$^{2,3}$, Bryan Webber$^{2}$\\
$^1$CERN PH-TH, Geneva 23, 1211 Switzerland \\
$^2$Cavendish Laboratory, University of Cambridge, Cambridge, UK\\
$^3$Department of Applied Mathematics and Theoretical Physics, University of Cambridge, UK\\
Email: \email{gripaios@cern.ch, sakurai@hep.phy.cam.ac.uk, webber@hep.phy.cam.ac.uk}
}

\abstract{
We consider the problem of reconstructing energies, momenta, and masses in collider events with missing energy, along with the complications introduced by combinatorial ambiguities and measurement errors. Typically, one reconstructs more than one value and we show how the wrong values may be correlated with the right ones. The problem has a natural formulation in terms of the theory of Riemann surfaces. We discuss examples including top quark decays in the Standard Model (relevant for top quark mass measurements and tests of spin correlation), cascade decays in models of new physics containing dark matter candidates, decays of third-generation leptoquarks in composite models of electroweak symmetry breaking, and Higgs boson decay into two $\tau$ leptons.
}   


\begin{document}   
%%%%%%%%%%%%%%%%%%%%%%%%%%%%%%%%%%%%%%
\section{Introduction}
%%%%%%%%%%%%%%%%%%%%%%%%%%%%%%%%%%%%%%
One of our cherished hopes for the LHC is that it will discover an elementary particle that constitutes the dark matter permeating our Universe.  Such a particle would necessarily carry neither electric nor colour charge and would be invisible in detectors, its presence being inferred from an excess of events with measured missing energy.

Though the discovery of a new invisible particle at the LHC would surely be serendipitous in itself, making the subsequent connection between such a particle and the dark matter in the cosmos presents a formidable challenge. To do so, one would need to measure the basic properties of the particle, such as its mass, spin, and couplings, in the laboratory setting. Such measurements are inevitably complicated by the fact that the particle, along with information about the energy and momentum that it carries, is lost.

The fact that information is lost does not render the situation hopeless, however. Indeed, in doing experimental data analysis, one is often faced with the situation that parameters are either poorly measured, or not measured at all: the remedy is simply to marginalize with respect to such parameters when computing the likelihood of some hypothesis. But in order to do so, one needs to have a well-defined hypothesis. This strategy has worked rather well for Standard Model physics, where, for example, the presence of invisible neutrinos in leptonic decays of the top quark has not prevented us from measuring the mass of the latter in that channel. But it is not obvious that the strategy will work well when it comes to new physics, beyond the Standard Model. One can, of course, simply impose a hypothesis by {\em fiat}, 
in the form of an explicit Lagrangian,
but then one runs the risk that the hypothesis may turn out be wrong, in which case inferences based upon it should not be trusted.

Another strategy is to hope that  Nature is benevolent enough to supply us with new physics, within the reach of the LHC, that allows us to make measurements whilst only assuming a much more general hypothesis.\footnote{Even with a more general hypothesis, a likelihood based analysis may bear fruit \cite{Allanach:2004ub,Webber:2009vm}.}  Preferably, one would like to make the hypothesis minimal, for example
assuming that new particles are produced in pairs and that each decays to the same
invisible particle.\footnote{Alternatively, one could add limited dynamic assumptions, in the form of an effective Lagrangian or ``simplified model'' \cite{Alves:2011wf}.}  This assumption typically holds in models where
dark matter is stabilized by a parity symmetry, e.g. supersymmety
with R-parity and universal extra dimensions with KK-parity. With this assumption, we can perform mass measurements purely based on kinematics [4.5];\footnote{For a review of kinematic methods for mass determination, see \cite{Barr:2010zj}.}
for other measurements, one might need to go further and reconstruct energies and momenta event-by-event. Examples discussed previously include measurements of spin \cite{Cheng:2010yy} and $CP$-violation \cite{MoortgatPick:2009jy}.
%for other measurements, one would need, in general, to go further and reconstruct energies and momenta event-by-event.

One could imagine doing so in a theory that predicts many new  particles, with masses roughly degenerate.
In such a theory, we expect that heavier new particles, once produced, will decay via a series of cascade two-body decays (which have greater phase space available than decays with three or more bodies, at least in the limit that the masses of the decay products may be neglected), terminating with the neutral, dark matter candidate particle.
%\footnote{An obvious candidate for such a theory is a supersymmetric extension of the Standard Model, with supersymmetry broken at a high scale. Such a theory certainly contains many new particles (at least one for every particle in the Standard Model), and a rough degeneracy in their spectrum is an inevitable consequence of renormalization group evolution from the SUSY-breaking scale down to the weak scale: even if some superpartner masses are very different from others at the high scale, the coupled evolution will pull the masses together. Several other scenarios for new physics predict similar sequential decay chains.}
%\footnote{An obvious candidate for such a theory is a supersymmetric extension of the Standard Model, with supersymmetry broken at a high scale. Such a theory certainly contains many new particles (at least one for every particle in the Standard Model), and their near degeneracy is an inevitable consequence of renormalization group evolution from the SUSY-breaking scale down to the weak scale. Several other scenarios for new physics predict similar sequential decay chains.}   
Each decay along the chain imposes a mass-shell constraint on the kinematics of the event. If there are enough constraints, all energies and momenta may be reconstructed. However, one still faces a number of difficulties in reconstructing such events at the LHC:
\begin{enumerate}[(1)]
\item The new particles result from collisions between constituent partons of the incoming protons, whose momenta are unknown.
\item By definition, the processes of interest involve invisible final-state particles.
\item Some of the decay products may be coloured partons, which manifest themselves in the detector as hadronic jets.  There are intrinsic uncertainties in reconstructing the kinematics of the parent partons from their associated jets --  the jet energy and angle resolution of the detector, the treatment of jet masses, hadronization and underlying event effects, etc.
\item There are often combinatorial ambiguities in assigning final-state objects to the decay chains.
\end{enumerate}
In an ideal world with only difficulties (1) and (2), mass-shell constraints plus missing transverse momentum measurements (for single events or multiple events of the same process) can suffice  to reconstruct full event kinematics.  However, even in this case equations for unknown masses or momentum components are typically polynomials, with multiple solutions, only one of which is correct.  The question then arises: how should we choose between the solutions?

These polynomials have real coefficients so their roots must be real or else complex-conjugate pairs. Often the polynomials are of even degree, in which case there is at least one incorrect real root accompanying the correct one.

Some of the polynomial roots do not correspond to solutions of the original kinematics.  This is obviously the case for complex roots, but may also arise for real roots.
There are two reasons for this. One is the logical abhorrence, familiar to all of us from our schooldays, that, while $x = y$ implies $x^2 = y^2$, $x^2 = y^2$ implies either $x = y$ or $x= -y$. Thus, while a root of the constraint equations is also a root of a polynomial equation that is obtained from them by a process of squaring operations, the converse is not necessarily true. The other reason is that roots may be physically unacceptable on other grounds, for example if a reconstructed energy exceeds the centre of mass energy available in a collision.

But often there are multiple acceptable real roots.  In any case, in the presence of the uncertainties (3) the correct root could become complex.  These effects can be regarded as perturbations of the coefficients of the polynomial.  Then the only way the correct root can become complex is if, as the perturbation is increased, it collides with an incorrect real root and they both move into the complex plane in conjugate directions.  This requires that the correct and incorrect roots are close together in the absence of the perturbation. 

Remarkably, it often happens that correct and incorrect real roots are indeed close together.  This can happen if the process involves a sequential decay chain with a large mass hierarchy, or conversely an approximate degeneracy.
To be explicit, consider the problem of reconstructing the mass of some particle at the head of a cascade decay chain.

(a) If there is a large hierarchy, a mass in the chain is approximately zero on the scale of masses higher up the chain.  If $p$ and $q$ are the 4-momenta of decay products of a zero-mass object, which must themselves have zero mass, the mass-shell condition $(p+q)^2=0$  implies $p\propto q$, which represents two more constraints than $(p+q)^2=m^2$, so the number of solutions is reduced.  This means that roots of the polynomial must coalesce (or move to infinity, but then they cannot be genuine solutions of the kinematics).  So for an approximately zero mass, there may be an incorrect real root ``close to" the correct root.

(b) Similarly if the decay product with 4-momentum $q$ has mass equal to the parent mass $m$, then $p$ must be infinitely soft and the parent must also have 4-momentum $q$, which represents additional constraints.  So again roots must coalesce in this limit, and be ``close" near this limit. 

In fact, since correct and incorrect solutions must be perfectly correlated (in that they coincide) at both extremities of the range of possible intermediate mass values, it turns out that there is a high degree of correlation between correct and incorrect solutions for any values of the intermediate masses.

We shall see however that the ``closeness" of solutions (or, equivalently, their degree of correlation) is difficult to define quantitatively.  For certain kinematic configurations, divergence from the limit can be very rapid as the hierarchy or degeneracy is broken.  Nevertheless it means that in these circumstances even the incorrect roots will be more densely distributed near the correct value.  And in the presence of effects (3) the real parts of complex roots will be also tend to be close to the correct value, with small imaginary parts.  Therefore it can make sense simply to plot the real values of all solutions, with a consequent gain in statistics.

In the presence of combinatorial ambiguities (4), we cannot in general expect to get any real roots from wrong combinations.  The only general feature is that the polynomial coefficients are still real and so the roots must be real or occur in complex-conjugate pairs. However, if there is a hierarchy or near degeneracy there will be approximate permutation symmetries that imply that the corresponding wrong combinations have roots close to those of the right combination:

(a) When there is a mass hierarchy, visible objects further down the chain are approximately collinear and therefore permuting their momenta will not significantly affect the reconstruction of kinematics higher up the chain.

(b) When there is an approximate degeneracy, some visible momenta will be soft and permutation of these will also not significantly affect reconstruction.

As before, the correlation between right and wrong combinations, which is perfect at either end of the interval of allowed intermediate masses, persists throughout the interval of intermediate masses. 
Thus again in these cases it can make sense to take the real parts of all solutions for all combinations.  There will be a peaking around the true solution when the combination is right, and also when the combination corresponds to an approximate permutation symmetry, plus a ``background" due to non-symmetric wrong combinations.

In a later section, we shall present an abstract discussion of these phenomena, showing that they have a natural formulation in terms of the theory of Riemann surfaces. %This section may be skipped by readers who find themselves mathematically indisposed. 
We shall also investigate, via a combination of analysis and numerical simulations, several examples. Before doing that, we would like to whet the reader's appetite by means of an illustrative example, which is not only simple enough that the behaviour may be understood without too much effort, but also is relevant for collider physics today. The example concerns measuring the mass of a top quark decaying in the leptonic channel.

As this example makes clear, our insights are not limited to new physics, beyond the Standard Model. Indeed, experimental analyses involving event reconstruction techniques are ubiquitous in collider physics. As an example, whenever one observes a lepton in association with missing energy at a hadron collider, one has the option of using the known mass of the $W$ boson to reconstruct the four-momentum of a hypothesized $W$-boson in the event. This information might then be used to study, for example, spin correlations or asymmetries (charge or forward-backward) in pair production of top quarks, or to reconstruct a resonance in the $WW$ channel (such as the Higgs). Similarly, whenever one observes a $\tau$ candidate, one may reconstruct the momentum of the $\tau$ by assuming that the neutrino emitted in the $\tau$ decay is collinear with the visible decay products \cite{Ellis:1987xu}. Until recently, this method was employed by both the ATLAS \cite{Aad:2009wy,ATLAS1} and CMS \cite{CMS1} collaborations in their strategies for searching for Higgs bosons. However, requiring that the reconstructed momenta be physical forces one to discard up to half of the events \cite{Aad:2009wy,ATLAS1}, in the presence of detector resolution, and this strategy has been abandoned in recent studies \cite{ATLAS2,CMS2}.
We present a different method for reconstructing events, using the information that comes from the secondary vertex in $\tau$ decays. We argue that in this case it makes sense to retain unphysical solutions, with a consequent gain in statistics.

Moreover, even these examples involving SM neutrinos have applications in new physics searches: reconstruction of the $W$ mass in this way was used recently to look for a resonance in the dijet plus $W$ channel that might explain the recent anomalous excess observed by CDF \cite{Aaltonen:2011mk}. It has also been suggested as a way to discover (and distinguish between) a new $tt$ or $t\overline{t}$ resonance in the di-leptonic channel \cite{Bai:2008sk}.
%%%%%%%%%%%%%%%%%%%%%%%%%%%%%%%%%%%%%%
\subsection{The top quark example}\label{sec:top}
%%%%%%%%%%%%%%%%%%%%%%%%%%%%%%%%%%%%%%
Consider a top quark, $t$, decaying to a bottom quark, $b$ and a $W$-boson, which in turn decays to a lepton, $l$ and an invisible neutrino, $\nu$ in $3+1$ spacetime dimensions, with the neutrino momentum in the two directions transverse to the beam inferred from the missing transverse momentum in the event.\footnote{For pair produced top quarks, we assume that the other top quark decays to visible hadrons.} We denote the mass and four momentum of particle $i$ by $m_i$ and $p_i^\mu = (E_i, \mathbf{p}_i,q_i)$ where $\mathbf{p}$ are the momentum components in the two directions transverse to the beam. The mass shell constraints then read
\def\vp{\mathbf{p}}
\def\svp{\slashed{\vp}}
\begin{align}
m_t^2 &= (p_\nu + p_l + p_b)^2, \\
m_W^2 &= (p_\nu + p_l)^2,\\
m_\nu^2 &= p_\nu^2, \\
\vp_\nu &= \svp,
\end{align}
where we have enforced conservation of four-momentum and where $\svp$ is the inferred missing transverse momentum. Now, assuming the masses other than $m_t$ are already known, these constitute five equations in five unknowns, namely $ p_\nu$ and $m_t$. Thus one can hope to  reconstruct both the top mass and all the particles' four-momenta in an event.

A little algebra shows that the last four equations can be reduced to a quadratic equation in either the energy or longitudinal momentum of the neutrino. 
Hence, using the first equation, one may obtain a quadratic equation in $m_t^2$, with two real solutions, one of which must have the correct value of $m_t^2$.  Neglecting the masses of the $b$ quark, the lepton and the neutrino, the difference between the two solutions is given by
\begin{align}
\label{eq:top}
E_l\Delta E_\nu &= q_l\Delta q_\nu = \frac{E_l q_l}{\vp_l^2}\sqrt{(m_W^2 + 2\vp_l \cdot \svp)^2 - 4 \vp_l^2 \svp^2 },\\
\Delta m_t^2 &= 2(E_b \Delta E_\nu - q_b \Delta q_\nu).
\end{align}

This simple expression for the difference between the correct and incorrect solutions for the top mass already contains much information. Firstly, we see that, as the mass of the $W$ increases towards $m_t$, such that the $b$ becomes soft, the difference between correct and incorrect solutions for $m_t$ (though not for $E_\nu $ and $q_\nu$) vanishes. Secondly, we see that the differences all vanish as the mass of the $W$ decreases to zero, since in this limit the lepton and neutrino become collinear, such that  $(\vp_l \cdot \svp)^2 - \vp_l^2 \svp^2 \rightarrow 0$. Therefore we expect that near either of these limits, wrong solutions for the top mass will be densely distributed over many events near the right solutions. Thirdly, we see that if one starts at large enough values of $m_W$ (near $m_t$) and decreases $m_W$, the wrong solution will always begin by moving away from the right solution, eventually turning around and coming back towards it at small $m_W$. The turnaround point depends on the kinematics of a particular event, but it tells us that, at a fixed, small value of $m_W$ but with multiple, random events, we can expect that the wrong solutions will 
still be more densely distributed near the right ones. Nevertheless, sometimes the wrong solution will be rather far away from the right solution, leading to large tails in our distributions. Indeed, for the extreme case of events that have $\vp_l = \mathbf{0}$, we see that the wrong solution lies infinitely far away from the right solution. These events form a set of measure zero, but nevertheless, we learn that very large tails can arise.

We shall return to this example in Section \ref{sec:single}, where we shall provide a simple geometric explanation of the above phenomena and identify further interesting properties of the solutions. 

%%%%%%%%%%%%%%%%%%%%%%%%%%%%%%%%%%%%%%
\section{Generalities and connection with Riemann surfaces}
%%%%%%%%%%%%%%%%%%%%%%%%%%%%%%%%%%%%%%
In this section, we give a more abstract discussion
which, although (we hope) illuminating, is not necessary to understand the examples given in later Sections and may be skipped by readers who wish to avoid mathematical niceties.

Let us consider, then, some cascade decay or decays, in which the unknowns, corresponding variously to energies or momenta that go unmeasured (for example those of invisible particles such as neutrinos or dark matter candidates) or {\em a priori} unknown masses, are equalled or outnumbered by the constraints, coming from the mass-shell conditions and measurements of total ``missing'' momenta, inferred from global momentum conservation in an event. 
For the time being, we assume that there are no combinatorial ambiguities and that all quantities are well-measured.
This set of equations then has at least one solution (the right solution), but may also possess wrong solutions, which for a generic event will lie in a finite set.

As we have already remarked, there may exist limits of the parameters in which the number of constraints is effectively increased.
Now, it may be the case that these extra constraints are redundant, in the sense that they are already implied by the other constraints on the system. If they are not,
then the number of solutions will be reduced.

This reduction in the number of solutions begs the question: what happens to the other solutions as the limit is taken? In particular, where do the other solutions lie when one is close to the limit? Two possibilities suggest themselves. One is that the wrong solutions become larger and larger and eventually go to infinity. The other possibility is that multiple solutions coalesce in the limit, such that the differences between solutions are small close to the limit. If this is the case, then we have an effect whereby wrong solutions may lie close to right solutions, leading to an apparent correlation between the two in samples of multiple events.

Unfortunately, it is rather difficult to see explicitly from these generic arguments which of the two qualitative possibilities obtains; nor is it easy to decipher quantitatively, simply by staring at the system of constraints, how the number of solutions changes. To do that, it is convenient to reduce the set of constraints to a single equation in a single variable. Since the constraints involve, at worst, the square root operation, one can, by repeated squaring operations, always write this single equation as a polynomial equation in the single unknown, for which we wish to solve.  In what follows, we would like to study the behaviour of the solutions (or roots) of this unknown as another parameter in the system (an intermediate input mass, say) is varied. We can, by further squarings, always write the single equation as a polynomial in this parameter too, such that we arrive at a polynomial equation in two variables. This naturally leads us into a discussion of Riemann surfaces.  

Before that, we remind the reader that the process of squaring operations just described introduces an unpleasant complication: solutions of the polynomial need not be solutions of the original constraint equations.  We shall see explicitly that this can happen in one of our later examples. One should always check explicitly that solutions obtained from the polynomial are indeed {\em bona fide} solutions of the original system of multiple equations.
%%%%%%%%%%%%%%%%%%%%%%%%%%%%%%%%%%%%%%
\subsection{Connection with Riemann surfaces\label{sec:riemann}}
%%%%%%%%%%%%%%%%%%%%%%%%%%%%%%%%%%%%%%
Let us now consider our polynomial equation in two variables: one, say, an unknown mass $w$ (we choose the notation for this Section to match that of complex variable theory), and the other, say, an input mass $z$ of an intermediate particle somewhere further down the chain.
We seek the values of $w$, possibly complex, that result from real input values of  the known mass $z$. But the discussion will be clearer if we allow both to take complex values. So we have a polynomial, $P(w,z)=0$ of degree $(n,m)$, say. Ultimately, we wish to solve this for $w$ given some input value $z$, but for now, let us just consider it as a polynomial in two variables (or, an algebraic curve). 

Since this is an analytic constraint on two complex variables, it manifestly defines a Riemann surface, viz, a 1-complex-dimensional, analytic, manifold, $\mathcal{M}^g$, of genus $g$.\footnote{For a generic $P$, there exists a beautiful way to compute the genus of $\mathcal{M}^g$ directly from $P$ using the Newton polytope; sadly, we shall not need it here.}

We may also find two less explicit descriptions of the Riemann surface by solving $P(w,z) = 0$ to obtain two ``functions'' $w(z)$ and $z(w)$. These are, of course, multivalued, and have branch point singularities whenever the corresponding derivatives, $dw/dz$ or $dz/dw$, do not exist. Since $P$ is just a polynomial,
and since $$P = 0 \implies \frac{\partial P}{\partial w} \frac{dw}{dz} + \frac{\partial P}{\partial z}= 0,$$
the derivative $dw/dz$ exists unless $\frac{\partial P}{\partial w}$ vanishes. One can easily show, furthermore that this is the condition for the polynomial $P$, considered as a polynomial in $w$, to have a repeated root at some value of $z$.
The branch points of these functions then define a Riemann surface in the usual way: one makes arbitrary branch cuts, lifts the complex plane to a multi-sheeted cover and obtains a single-valued function on $\mathcal{M}^g$. 
 It is important to stress, however, that these two descriptions of the same Riemann surface (one arising from branch points of $w(z)$ and one from $z(w)$) are quite different. 
 Indeed, one is an $n$-sheeted cover and the other is $m$-sheeted. Moreover, their branch points are not the same. 
 
 Now, we are interested in the problem of finding the solutions for the unknown mass $w$ that result as we vary the input mass parameter $z$. The description of $\mathcal{M}^g$ that is relevant for us is therefore the one provided by the function $w(z)$. (If we were interested in the inverse problem of solving for $z$ given $w$, the appropriate description would be in terms of $z(w)$; we repeat that these two descriptions differ in their branch structure.)
 
 We may now ask what happens as we vary the input mass parameter $z$ along a trajectory in $\mathbb{C}$ that goes along the real axis from some initial value towards the origin, where the nature of the mass-shell constraint changes, such that the number of constraints increases. We already know that the behaviour of the solutions must be completely smooth, except for possible branch point singularities. We also expect that the number of solutions must decrease at the origin. We now ask what this implies for the Riemann surface. There are three possibilities, which we discuss in turn.

One possibility is that, due to the logical abhorrence mentioned above, some of the solutions of the polynomial simply cease to become solutions of the system of multiple equations. We shall see it explicitly in the examples. 
 
 The second possibility is that some of the roots go towards the point at infinity. Whilst perfectly acceptable from the point of view of the compact Riemann surface, we would no longer regard these as physical solutions. In our examples, this only happens for special kinematic configurations.
 
The third possibility, which is of most interest to us, is that the polynomial has a repeated root, or equivalently, that $w(z)$ has a branch point, at the origin in $z$. If so, in the neighbourhood of the branch point, multiple solutions will lie close together, leading to a correlation between correct and incorrect solutions, if one of those solutions is the correct solution. 

We note that the trajectories followed by the roots as the input parameter moves towards the branch point at the origin may be highly non-trivial, as the reader may see by glancing ahead at Figures \ref{fig:2d}, \ref{fig:lqkW}, and \ref{fig:lqkT}, which illustrate the later examples. The left-hand column of each Figure shows the trajectories, projected from $\mathcal{M}^g$ into the complex plane, followed by the roots in an event. We shall discuss these in more detail later. For now, we note that the roots can indeed coalesce at branch points, that they can move away from the branch point before moving towards it, and also that they can reverse, or otherwise change, their direction, following a cusped trajectory. The cusps do not correspond to singular branch points of the description of the Riemann surface in terms of $w(z)$, which, as we discussed above, arise when $dz/dw$ vanishes (and are forced to lie on the real axis in the projected $w$-plane, given that the coefficients of $P(w,z)$ are real and that we follow a real trajectory in $z$). Rather, they arise at the branch points of the dual description of $\mathcal{M}^g$ in terms of $z(w)$, where $dw/dz$ vanishes. Indeed, at such points, then writing $w$, $z$ in terms of their real and imaginary parts, $w = u + iv$, $z = t$, we have that  
 $d u/d t= d v/d t = 0$, whence $d v/d u$ is undefined. A classic example is the cycloid curve, $u= t - \sin t, v =1 - \cos t$, which despite being a smooth map from $t$ to $(u,v)$ has cusps at the points where $d u / d t = 0$ and $dv/du$ is undefined.
%%%%%%%%%%%%%%%%%%%%%%%%%%%%%%%%%%%%%%
\subsection{Combinatorics}
%%%%%%%%%%%%%%%%%%%%%%%%%%%%%%%%%%%%%%
The limit as one of the intermediate masses goes to zero is also interesting from the point of view of the problem of combinatorial ambiguities. There are two types of combinatorial ambiguities. One arises when different visible particles along a decay chain are indistinguishable in particle detectors. The second arises when 
new particles are pair produced, and the subsequent decays involve identical (or rather indistinguishable) final states. In particular, if the branching ratio for one decay dominates over all others, then the decay chains will be identical (modulo charge conjugation), leading to an ambiguity in assigning observed final state particles to one or other decay chain.

If such ambiguities are truly ambiguous, then the only robust manner in which to proceed is to consider all possible assignments in solving the system of constraints. For a $p$-fold ambiguity, one must solve the constraint system $p$ times, obtaining $p$ copies of all solutions, both right and wrong. Of course, only one of these solutions is the correct one.

Now, in the limit that an intermediate mass goes to zero, it is easy to see that ambiguities in the arrangement of visible particles further down the chain are irrelevant, in the sense that the solutions of the constrained system after permutation are the same as the original solutions. Why? In the limit that an intermediate mass goes to zero, all subsequent particles (which must also be massless) must be emitted collinearly. They may be fully characterized by the fraction of the energy of the parent particle that they carry, such that the order of emissions is irrelevant.

Since permutations down the chain are irrelevant in the limit that the mass vanishes, and since we expect smooth behaviour in the solutions as the mass varies (for a wrong permutation, we are simply solving a different polynomial, and we still have a Riemann surface, albeit a different one), then for small intermediate masses, we should find that solutions of the permuted equations are close to right or wrong solutions of the equations with the correct particle assignment, for which the wrong solutions may, in turn, be close to the right solution.

We now pause to remark that there is an important distinction between the reality properties of solutions (right and wrong) of the right equations and those of the wrong equations, {\em viz.} those obtained by a wrong permutation. In the former case (in the absence of measurement errors), we are guaranteed that one of the solutions (the right one) is real. (For an even polynomial, we are also guaranteed that there exists another real solution, which may or may not lie close to the right solution.) When we solve the polynomial equation corresponding to a wrong particle assignment, we are not guaranteed any real solutions. Nevertheless, according to the arguments above, we expect solutions lying close to the real solution, but possibly off the real axis, in the limit than an intermediate mass is small.

Na\"{\i}vely therefore, we can reduce the combinatorial ambiguity by only accepting solutions that are real. As we shall now discuss, this is not necessarily the optimal strategy in the presence of measurement errors.
%%%%%%%%%%%%%%%%%%%%%%%%%%%%%%%%%%%%%%
\subsection{Mismeasurements}
%%%%%%%%%%%%%%%%%%%%%%%%%%%%%%%%%%%%%%
In the presence of measurement errors, none of the solutions obtained is the right solution. Moreover, we are not even guaranteed to have any real solutions of our polynomial equation, even with the correct particle assignment. This then re-opens the question of whether one should insist on real solutions, as in \cite{Cheng:2008mg}, or whether one should accept all complex solutions, or only those whose imaginary part is small, according to some criterion.

Let us now consider this issue in more detail. At least in the presence of arbitrarily small measurement errors, it makes sense to retain only real solutions. Indeed, since we are solving real polynomials, the complex solutions may only occur in complex-conjugate pairs. Starting from the limit in which measurement errors vanish and one solution is the truly right solution, we see that this solution must remain real as we increase the measurement error, unless the measurement error is so large that the right solution can `pair up' with another solution and move off the axis. In order to do so, the error in the solution for the mass resulting from the measurement error must be comparable to the distance between the right solution and another wrong solution. If this distance were of order of the mass itself, then one could argue that one should reject complex solutions. Indeed, for such a solution to arise from the right, real solution would require a large measurement error, in which case the event should probably have been discarded in the first place.

Unfortunately, we have argued above that the distance between right and wrong mass solutions in the complex plane is not necessarily of the order of the mass itself.
On the contrary, we have argued that right and wrong solutions may coalesce in the limit that an intermediate mass becomes small. So it is easier than one might expect for the right, real solution to become complex in the presence of measurement errors. Of course, if one has a good understanding of the size of typical measurement errors, one could choose whether or not to accept complex solutions. In the absence of such an understanding, it would perhaps make more sense to accept all complex solutions. 
%%%%%%%%%%%%%%%%%%%%%%%%%%%%%%%%%%%%%%
\subsection{Classification of event reconstruction}
%%%%%%%%%%%%%%%%%%%%%%%%%%%%%%%%%%%%%%
Before discussing specific examples, let us attempt to categorize the different kinds of event reconstruction that one may envisage and give examples of them. 
We will show that several different kinds of reconstruction problem can be viewed as extensions of a basic momentum reconstruction problem.

The basic problem we consider is to reconstruct the energy and momentum of one or more invisible particles, in a single collision event, in which the masses of all particles are assumed to be known. One example relevant for colliders is the leptonic decay of a $W$-boson, where the neutrino has four unknown energy-momentum components, but there are four constraints, namely the two mass-shell constraints for the $W$ and the neutrino, and the two missing transverse momentum constraints. A second example is the di-leptonic decay of pair-produced top quarks. Here, there are two neutrinos and eight unknown energy-momentum components, but there are also eight constraints, if all the masses are known. 

Now consider a momentum reconstruction problem of this type, but in which there are more constraints than unknowns. Of course, one can still solve for the momenta if all the masses are known, but one can go further, since the constraints then imply relations between the masses of particles involved in an event. This then gives the possibility of reconstructing not only the momenta, but also some or all of the masses. Indeed, even with just one event, then if one already knows some of the masses, one may be able to solve for the others. With the masses known, one can then go back and reconstruct the momenta in that one event or indeed in any other. As a trivial example, one can always turn a momentum reconstruction problem into a mass reconstruction problem by adding one more particle (of unknown mass) at the head of a decay chain. As an example, taking the decay of a $W$-boson above, one may add a top quark that decays to it (together with a $b$) and solve for the mass of the top. This is precisely what we did in the introduction.

One may go even further: given that a single event of this type implies relations between the masses, one can attempt to reconstruct all of the masses by simply combining events. A possible collider example (which we shall study further later on) is given by pairs of cascade decays, each with three visible particles on each chain. There are eight unknown energy-momentum components (in 3+1 dimensions), but ten mass-shell and two missing transverse momentum conditions. Thus each event can be reduced to two relations on the particle masses. If the chains are assumed to be identical, such that there are only four independent masses in the chains, one needs two events to reconstruct all masses. If the chains are not identical, one needs four events.

In conclusion, we see that various mass reconstruction problems can be viewed as extensions of the basic momentum reconstruction problem.
%%%%%%%%%%%%%%%%%%%%%%%%%%%%%%%%%%%%%%
\section{Examples}
\label{sec:examples}
%%%%%%%%%%%%%%%%%%%%%%%%%%%%%%%%%%%%%%
%%%%%%%%%%%%%%%%%%%%%%%%%%%%%%%%%%%%%%
\subsection{Single chain decays \label{sec:single}}
%%%%%%%%%%%%%%%%%%%%%%%%%%%%%%%%%%%%%%
We have already given a algebraic discussion of one example, namely that of leptonic decays of the top quark.  We saw that, once the $W$-boson mass is known and the neutrino mass is assumed negligible, one may solve a quadratic equation for the mass of the top quark; this quadratic equation reduces to a linear equation in the limit that $m_W/m_t \rightarrow 0$ and this in turn leads to a correlation between the right and wrong solutions for small, but non-vanishing $m_W/m_t$.
We would now like to generalize this example further and show that its behaviour may be understood via simple, geometric arguments.

Consider a single decay chain in $D+1$ spacetime dimensions, $\dots \rightarrow C+ \dots \rightarrow B+2 + \dots \rightarrow A + 1+2+ \dots$, with visible particles $1,2,\dots$, terminating in an invisible particle $A$. We assume that the visible particles are all massless and that the masses of all states are known, and that we wish to solve for the unknown energy-momentum components of $A$. We assume that $0 \leq d \leq D$ of the spatial momentum components of $A$ can be inferred via some kind of missing energy measurement. By analogy with a collider physics experiment (and in a slight abuse of terminology), we will call these the `transverse' directions; the unmeasured momentum directions will be called `longitudinal'. We thus have $D+1-d$ unknowns and we may solve for these provided we have an equal (or greater) number of mass-shell constraints. We therefore need a chain containing (at least) $D-d$ visible particles. 

As it stands, this is a momentum reconstruction problem. We may turn it into a mass reconstruction problem by adding one more parent particle of unknown mass at the top of the chain. This adds one more unknown (the parent mass), together with one more mass-shell constraint, so the system remains constrained. There are then $D-d+1$ visible particles. In the example of the top quark decay described above, we have $D=3$, $d=2$, such that we need one visible particle (the lepton), to solve for the momentum of the neutrino, and two visible particles (the lepton and the bottom quark) to solve for the mass of the top.

The general constrained single cascade just described can be easily understood in a geometrical way, given the following lemma: provided we only consider Lorentz boosts in the subspace that is orthogonal to the transverse directions, then, if we consider two frames $F$ and $F^\prime$ related by such a Lorentz boost $\Lambda$, then the boost of a solution of the equations written in frame $F$ is itself a solution of the equations written in the boosted frame $F^\prime$. This is obviously true for the right solution, but the Lorentz invariance of the mass-shell constraints guarantees that it holds equally true for wrong solutions as well. An immediate corollary is that, if the unknown being solved for is a mass, the solution will be the same in any two such Lorentz frames.

We stress that the lemma does not hold if one considers boosts in the transverse directions, since there is no sense in which the missing-momentum constraints are Lorentz covariant. One does not even know how to boost the measured missing transverse momentum to another frame, since the result depends on the unknown missing energy.

\begin{figure}\begin{center}
\includegraphics[width=0.6\linewidth]{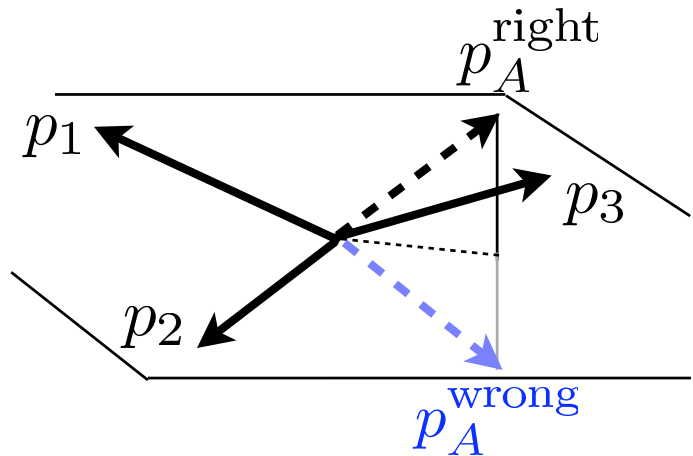}
\caption{The right and wrong solutions for the longitudinal momenta in the boosted frame, for a single chain decay with three visible particles.}
\label{fig:reflection}
\end{center}\end{figure}

With the lemma in hand, it is easy to see what happens. In general,
we can boost to a frame in which all the longitudinal momenta of the $D-d$ visible particles required for the momentum reconstruction problem
are linearly dependent, spanning a $D-d-1$-dimensional subspace. For example, we can boost to the longitudinal centre of mass frame of the $D-d$ visibles, in which the longitudinal momenta sum to zero. Figure \ref{fig:reflection} shows the particles in this frame in a decay with three visible particles. In this frame, the mass-shell constraints are invariant under a change in sign of the longitudinal momentum component of $A$ that is orthogonal to the subspace spanned by the visible longitudinal momenta. The two solutions of the equations ({\em viz.} the energy-momenta of $A$) are thus degenerate in this frame, with the exception of that orthogonal component, for which the two solutions are equal in magnitude, but opposite in sign. In a different frame, the longitudinal boost will of course mix up the components, such that none will be degenerate in general.
We note that this argument does not work in $D=1 \implies d=0$, because there is then only one (massless) visible particle, and no finite boost will take us to its rest frame. Indeed, explicit solution in that case shows that there is only ever one solution. More generally, whenever $D-d=1$, implying only one visible particle, the argument applies only if we can do a boost to the longitudinal rest frame of the visible particle. Since the particle is massless, we may do so only if the transverse momentum is non-vanishing.

Now, what happens when one of the intermediate masses is sent to zero? Then the energy-momentum of $A$ is necessarily collinear with the energy-momentum of visible $1$ in the above frame (and indeed in any other frame). Thus, the momentum component of $A$ orthogonal to the subspace spanned by the visibles is zero, and the two solutions for the energy-momentum of $A$ are degenerate in all components, in this frame, as are boosts thereof.

Turning now to the related mass measurement problem (with one more particle added to the chain), we see that, for all intermediate masses non-vanishing,
there will be two solutions for the mass of the added parent particle, obtained by plugging the two values for the energy-momentum of $A$ into the mass shell constraint for the parent particle. By the lemma, these two values will be the same in all frames related by longitudinal boosts. When an intermediate mass vanishes, the two values for the reconstructed energy-momenta of $A$ are the same, and so are the two values for the reconstructed parent mass. 
Finally, when any intermediate mass is small compared to the mass that preceeded it in the chain, the two reconstructed parent mass values should lie close together. 

These properties are all confirmed by an explicit algebraic analysis. In the momentum reconstruction problem, one obtains a quadratic equation (unless $D-d=1$ and the transverse momentum of particle $1$ is taken to zero, in which case the coefficient of the quadratic term goes to zero) which reduces to a linear equation in the limit that an intermediate mass vanishes.  For $D=3$ and $d=2$, one has a simple generalization of the top decay example considered in Section~\ref{sec:top}, viz.\ a single decay chain $C\to B+2$, $B\to A+1$ with an invisible particle $A$ carrying away the missing transverse momentum $\svp$.  In the case of top decay,  $A,B,C,1,2=\nu,W,t,l,b$.  Taking the masses of $A$ and $B$ as known and neglecting those of the visible decay products $1$ and $2$, we have in analogy with (\ref{eq:top}) 
\begin{align}
\label{eq:single}
E_1\Delta E_A &= q_1\Delta q_A = \frac{E_1 q_1}{\vp_1^2}\sqrt{(m_B^2 -m_A^2+ 2\vp_1\cdot \svp)^2 - 4 \vp_1^2 (\svp^2+m_A^2) },\\
\Delta m_C^2 &= 2(E_2 \Delta E_A - q_2 \Delta q_A).
\end{align}
Then for $m_B\to 0$ all the solution differences vanish since we must have $m_A\leq m_B$.   We also see that the two solutions for $m_C$, but not for $E_A$ and $q_A$, coincide when visible particle 2 is soft, corresponding to $m_C=m_B$.

Since the difference between right and wrong mass solutions vanishes when the intermediate mass takes its maximum and minimum values and is non-vanishing elsewhere, then the wrong solution (which is necessarily real) must change direction when we follow a trajectory that covers the full range of intermediate masses.
This illustrates in a rather extreme way one type of behaviour we described earlier in our discussion of Riemann surfaces: in this case, not only does the wrong root move away from the right root before returning, but the fact that it is also forced to be real means it traverses a cusp of angle $\pi$ as it does so. As discussed, this must correspond to a branch point of the dual description of the Riemann surface. 

\begin{figure}\begin{center}
\includegraphics[angle=90,width=0.8\linewidth]{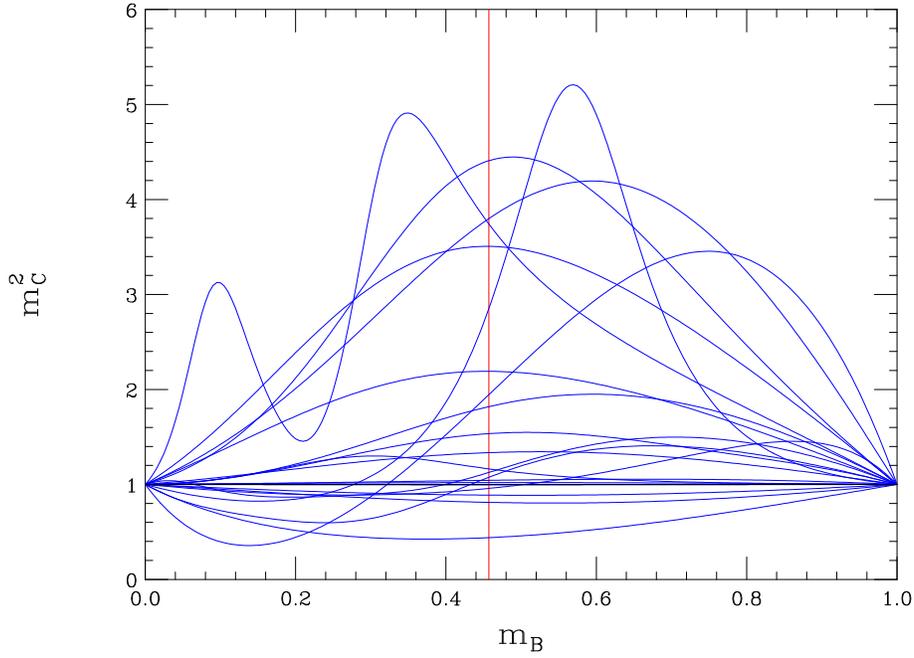}
\caption{Solutions for the mass-squared of the parent particle $C$ in the single decay chain $C\to B+2,\;B\to A+1$, as functions of $m_B$, for 20 ``typical'' events.  The correct solution $m_C=1$ is shown in black; the incorrect ones are in blue.  While $m_B$  is varied, the events have fixed decay angles in the parent ($B$ and $C$) rest frames, distributed isotropically.  The vertical red line corresponds to the kinematics of top quark decay.
\label{fig:single}}
\end{center}\end{figure}

Figure~\ref{fig:single} shows the trajectories followed by the wrong solutions as the intermediate mass $m_B$ is decreased from $m_C$ to zero, in a sample of twenty events. As expected, each trajectory begins and ends at the right solution ($m_C =1$), but departs from it in the intervening region. Moreover, whilst 
for the majority of events the wrong solution lies close to the right solution throughout the trajectory (including the red vertical line, which corresponds to the kinematics of top quark decays), the discrepancy can be large. Finally, we see that the trajectories can change direction more than once.
%%%%%%%%%%%%%%%%%%%%%%%%%%%%%%%%%%%%%%
%%%%%%%%%%%%%%%%%%%%%%%%%%%%%%%%%%%%%

There is yet one more interesting property of this decay chain, which is not evident in our geometrical description. We find that the difference in right and wrong solutions is independent of the mass of $A$, as one varies the mass of the intermediate $B$, whilst keeping the decay angles of all particles constant, as measured in their rest frames. This behaviour is easily demonstrated from Eq.~(\ref{eq:single}).  The momenta of particles 1 and $A$ in the rest frame of $B$ have magnitude $p^* = (m_B^2-m_A^2)/2m_B$.  Writing $\svp = \vp_B-\vp_1$, the argument of the square root in (\ref{eq:single}) can then be expressed as
\beq
4\left[(m_B p^* +\vp_B\cdot\vp_1)^2 -\vp_1^2(m_B^2+\vp_B^2)\right]\,.
\eeq
\def\vn{\mathbf{n}}
 Now as we are assuming particle 1 to be massless, its 4-momentum in the collider frame is of the form $p_1^\mu=p^*n^\mu$ where $n^\mu=(n^0,\vn,n^3)$ is a function of the 4-momentum of $B$ and the direction of 1 in $B$'s rest frame.  The important point is that $n^\mu$ is independent of $m_A$.  Then (\ref{eq:single}) can be written as
\beq
\Delta m_C^2 = \frac 4{\vn^2}(E_2 n^3 - q_2 n^0)\sqrt{(m_B  +\vp_B\cdot\vn)^2 -\vn^2(m_B^2+\vp_B^2)}\,,
\eeq
which is manifestly independent of $m_A$, as the $p^*$ dependence has cancelled.  It follows that, for given production and decay distributions of $C$ and $B$, the curves shown are really only functions of the ratio $m_B/m_C$.  Accordingly we have marked the point corresponding to top decay as $m_B = m_C m_W/m_t = 0.46$.

Turning this argument around, we can say that, if the mass ratio $m_B/m_C$ has been determined, then the distribution of the wrong solutions provides information on the decay angular distributions, independent of the mass of $A$.

\begin{figure}\begin{center}
\includegraphics[angle=270,width=0.9\linewidth]{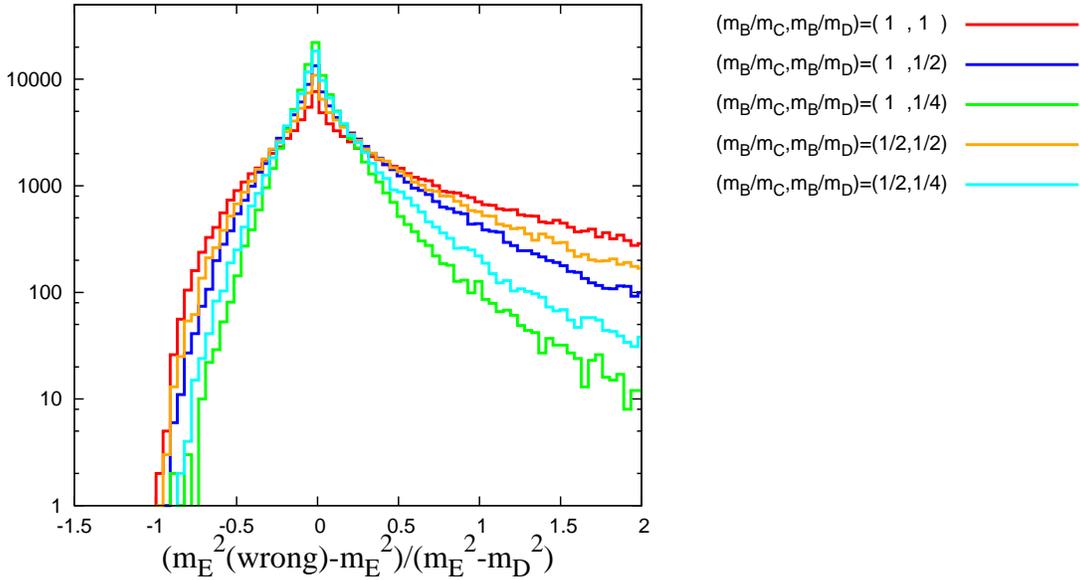}
\caption{
The distributions of $\Delta m^2_E /(m_E^2-m_D^2)$ for the decay chain
%$E \rightarrow D + 4 \rightarrow C +3 +4 \rightarrow B +2 +3+4 \rightarrow A + 1+2+3+4$,
$E \rightarrow D + 4, D\rightarrow C +3, C \rightarrow B +2, B\rightarrow A + 1$,
showing the correlation between right and wrong solutions.
}
\label{fig:mE}
\end{center}\end{figure}

Another interesting single decay chain is that with four visible particles,
$E\to D+4, D\to C+3, C\to B+2, B\to A+1$. 
In this case, given the masses of particles $A,B,C,D$, one obtains a quadratic equation for the mass of the parent particle $E$ without any missing momentum measurement.  The difference between the solutions takes the form
\beq
\Delta m_E^2 = (m_E^2 - m_D^2)f(m_B/m_C,m_B/m_D;\Omega)\,,
\eeq
where $\Omega$ represents the dependence on the decay angles.  Thus again the distribution of the wrong mass solutions is independent of the invisible particle mass $m_A$.  The function $f$ is complicated but vanishes as $m_C$ and/or $m_B\to 0$.  Therefore the solutions coalesce in these limits, and also as $m_D\to m_E$.
The fact that the wrong mass solution is forced to lie close to the right mass solution
in the limit of either large or small intermediate masses leads to a correlation in the distribution of wrong solutions and the right solution in a sample of multiple events. This is illustrated in Figure \ref{fig:mE}, which shows the distribution of the wrong mass solution obtained in a decay chain with four visible particles, for varying values of the intermediate masses.

Shorter single chain examples where no missing energy measurement is available may also be relevant for collider physics. Here one needs to combine information from multiple events (making the hypothesis that each corresponds to the same signal), in order to obtain a constrained system \cite{Nojiri:2003tu,Kawagoe:2004rz}. 
%%%%%%%%%%%%%%%%%%%%%%%%%%%%%%%%%%%%%%
\subsection{Double chain decays \label{sec:double}}
%%%%%%%%%%%%%%%%%%%%%%%%%%%%%%%%%%%%%%
Let us now turn to pair produced particles, each of which undergoes a cascade decay to an invisible particle. We label one chain as before, and the other with primes: $\dots \rightarrow C^\prime + \dots \rightarrow B^\prime +2^\prime + \dots \rightarrow A^\prime + 1^\prime +2^\prime + \dots$. It is not necessary to assume that the two decay chains are identical, or even of the same length.

Now, in the absence of measured missing energy, the constraints on the two cascade decays are decoupled from each other; we can, thus, apply independent Lorentz boosts to the two chains and show that, as above, solutions exhibit a pairwise degeneracy in the limit that intermediate masses vanish.

Even in the presence of missing energy constraints that couple the two chains, we may be able to reconstruct the two cascade decays individually, in which case the arguments 
of the previous section still go through. Let us consider double chains, with $n$ and $m$ visible particles, in $D+1$ spacetime dimensions with $d$ measured missing momenta. 
Assuming all masses are known, to solve for the momenta of the two invisible particles $A$ and $A^\prime$, we must have that $2D = d + n +m$. For example if $D=3,d=2,n=3,m=1$, we can first solve for the $n=3$ chain, ignoring the missing energy and then reconstruct the $m=1$ chain using the missing energy. Again, this may be converted into a mass reconstruction problem by adding two parent particles, at the top of each chain, or indeed one parent particle at the top of both chains.

Novel cases arise when we cannot decouple the two chains. The simplest example is $D=d=2, n=m=1$. Whilst this example is not obviously relevant for hadron collider physics, it nevertheless provides a useful illustration of what may happen in situations that are relevant for colliders, such as $D=3,d=2,n=m=2$.

This $D=d=2, n=m=1$ example can, by elimination, be reduced to a quartic equation for the invisible particle momenta, with four complex roots, of which either two or four must be real, in the absence of combinatorics and measurement errors. We solve the quartic equation numerically for several events corresponding to the topology with a single particle $C$ at the head of two identical decay chains. In the limit that the masses of $B$ and $B^\prime$ vanish, the system of constraints collapses to a linear equation.  Indeed, in each chain, the visible particle 1 or $1^\prime$ is forced to be collinear with the invisible particle $A$ or $A^\prime$, such that we have the equations $p_A = \alpha p_1$ and $p_{A^\prime} = \alpha^\prime p_{1^\prime}$, with $\alpha, \alpha^\prime$ unknown. Plugging these into the two transverse missing momentum constraints gives a unique solution for $\alpha, \alpha^\prime$ and hence for all the other unknowns. 

Thus, provided solutions do not move off to infinity or cease to become solutions as 
$m_{B,B^\prime} \rightarrow 0$, then all four solutions must coalesce at that point, with the three wrong solutions lying on top of the right solution.

A complication arises when we try to solve for the mass of the parent particle $C$ at the head of the two chains.  This quantity involves the energies of the invisible particles $A,A'$ as well as their 3-momenta.  For the real solutions we can legitimately demand that these energies be positive, but for the complex solutions we have to accept either sign, making four mass solutions for each solution of the quartic equation.

\begin{figure}[ht!]\begin{center}
\includegraphics[angle=90,width=0.399\linewidth]{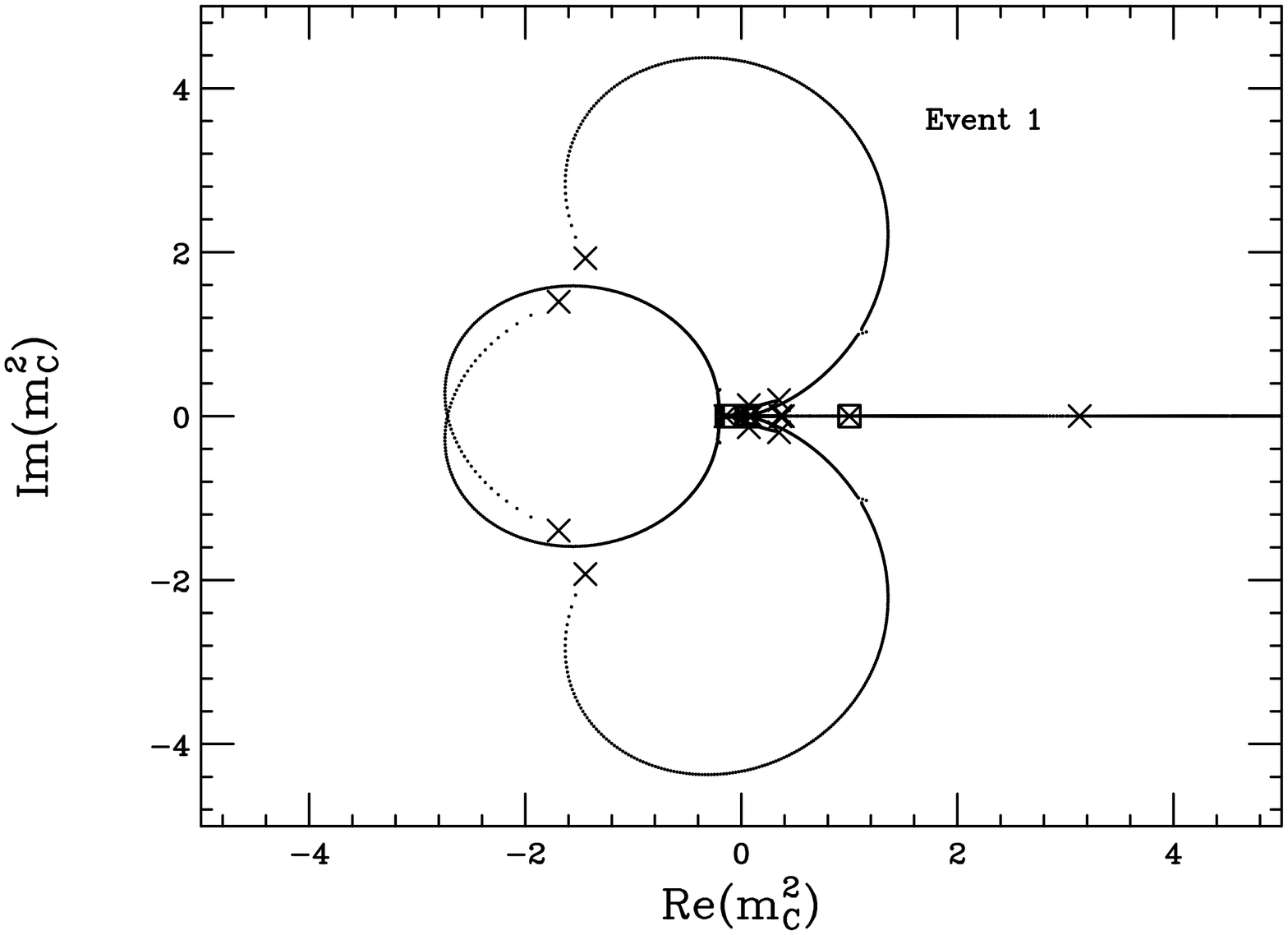}
\includegraphics[angle=90,width=0.399\linewidth]{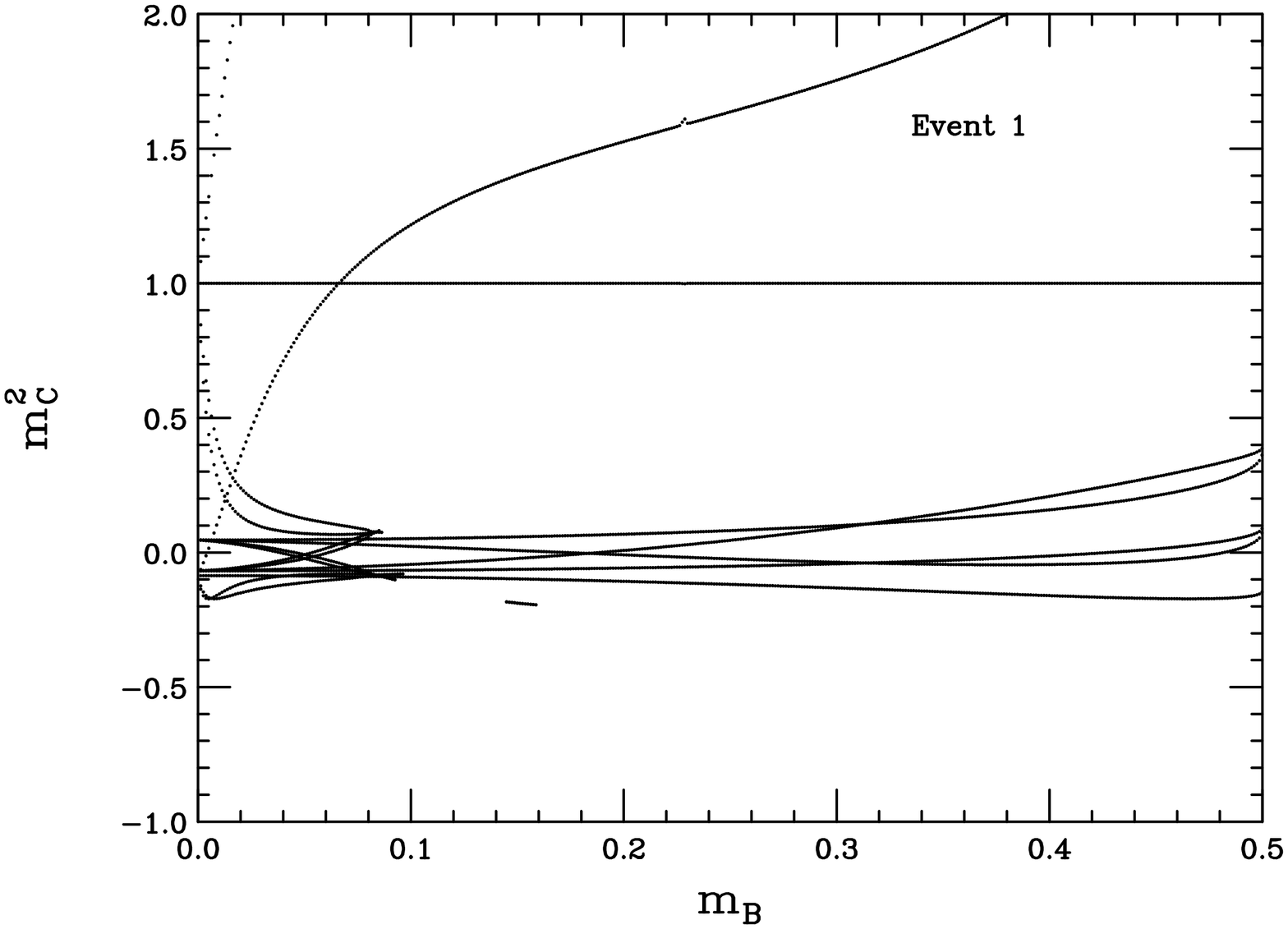}
\includegraphics[angle=90,width=0.399\linewidth]{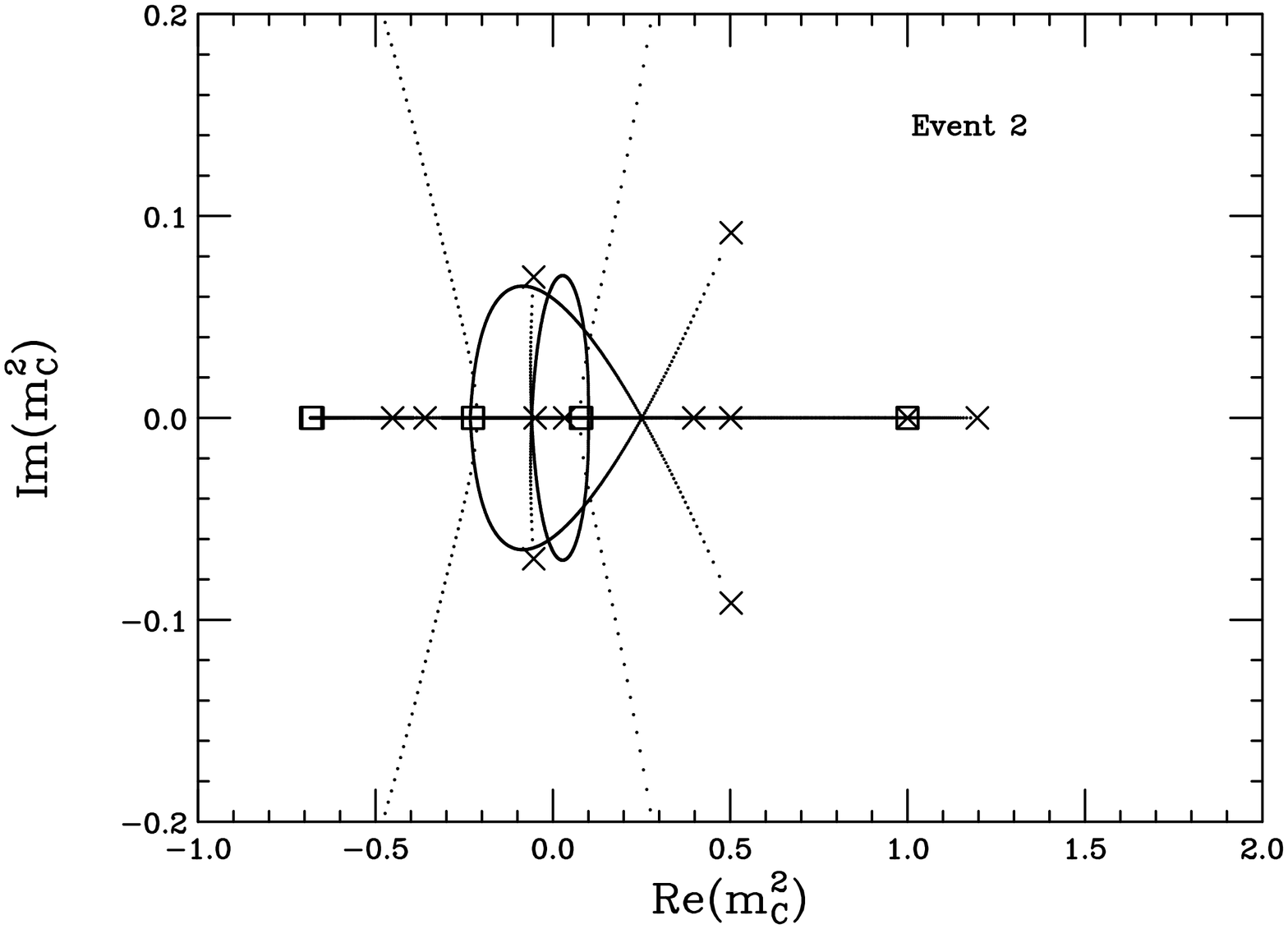}
\includegraphics[angle=90,width=0.399\linewidth]{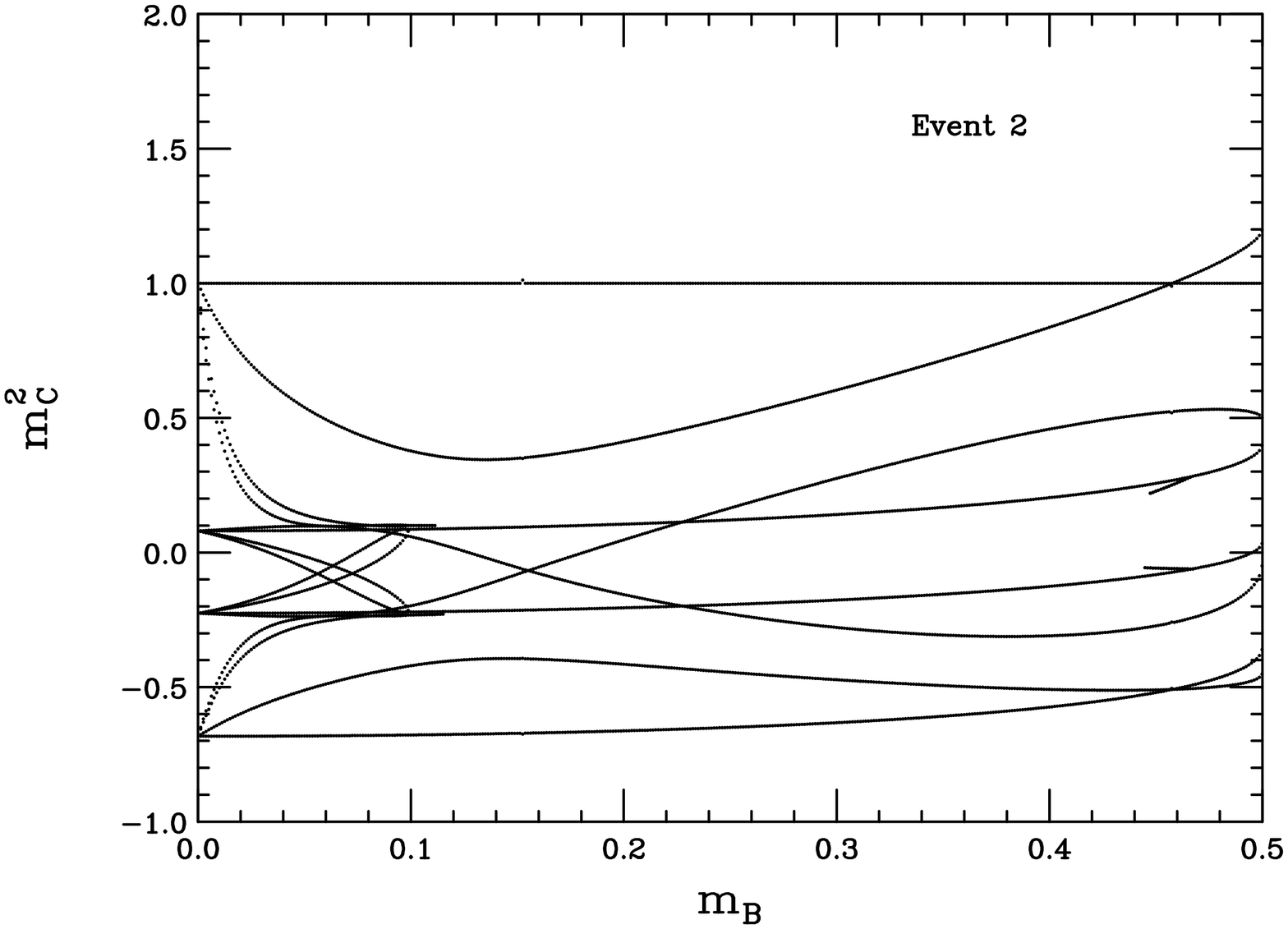}
\includegraphics[angle=90,width=0.399\linewidth]{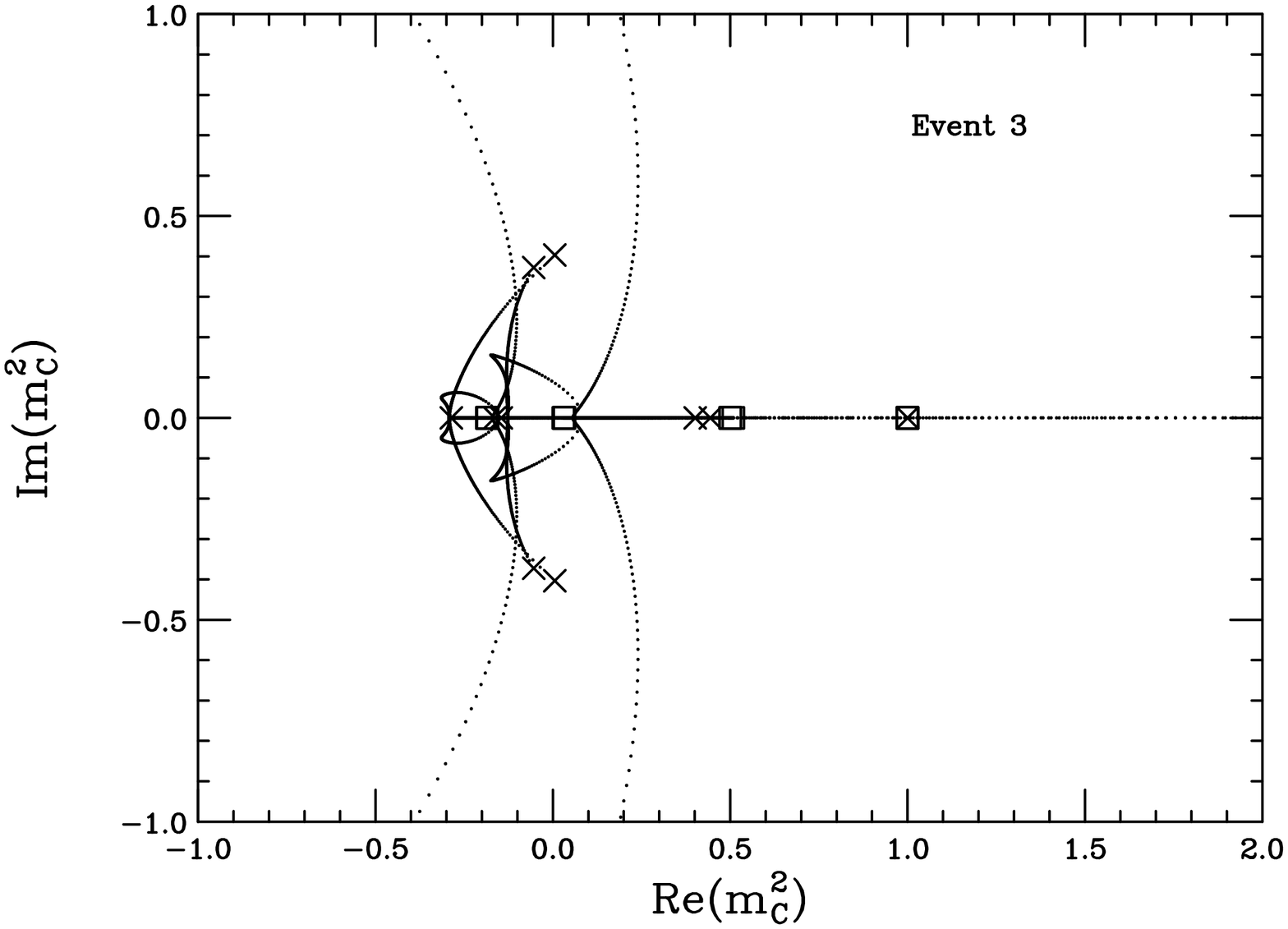}
\includegraphics[angle=90,width=0.399\linewidth]{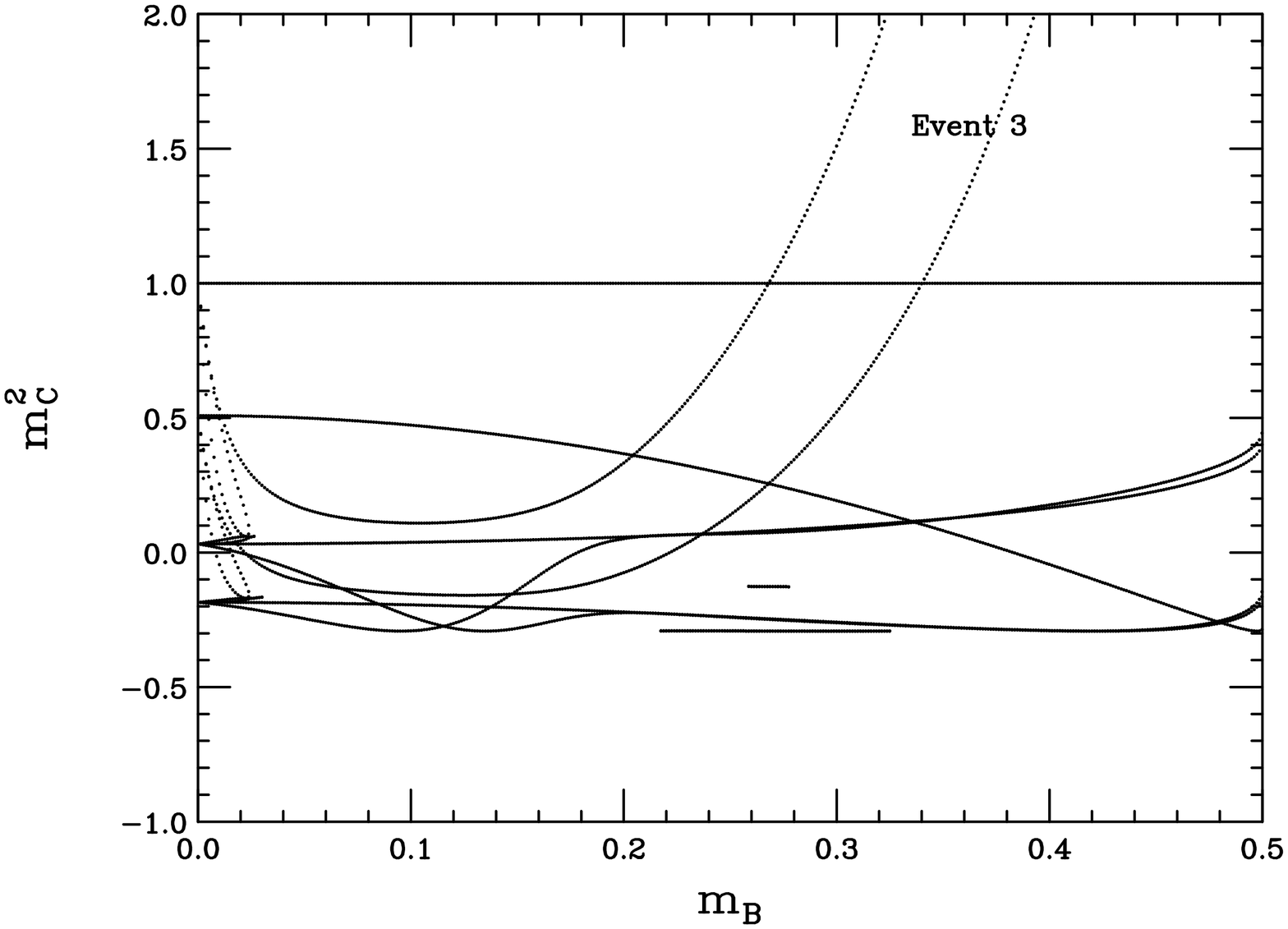}
\includegraphics[angle=90,width=0.399\linewidth]{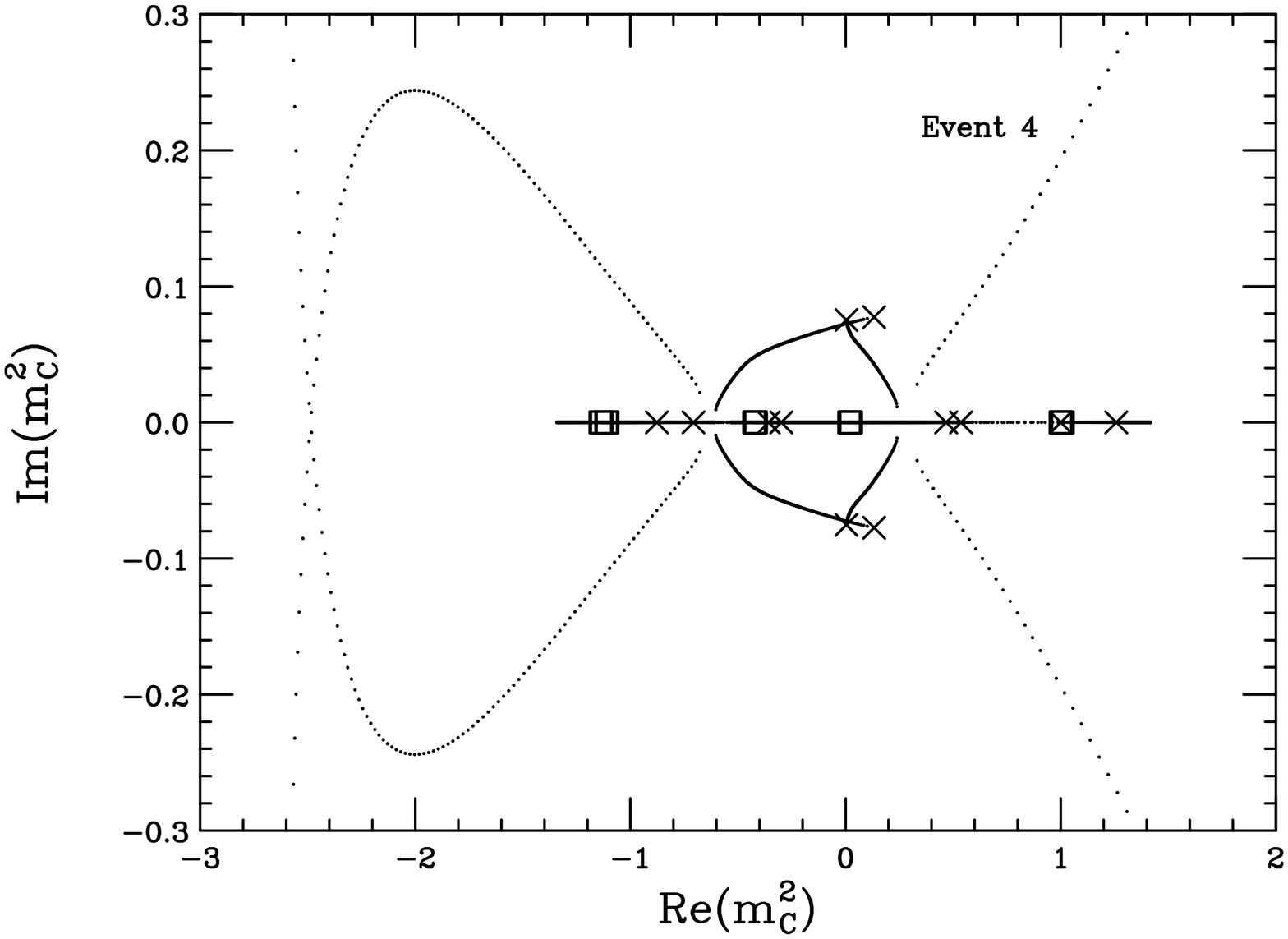}
\includegraphics[angle=90,width=0.399\linewidth]{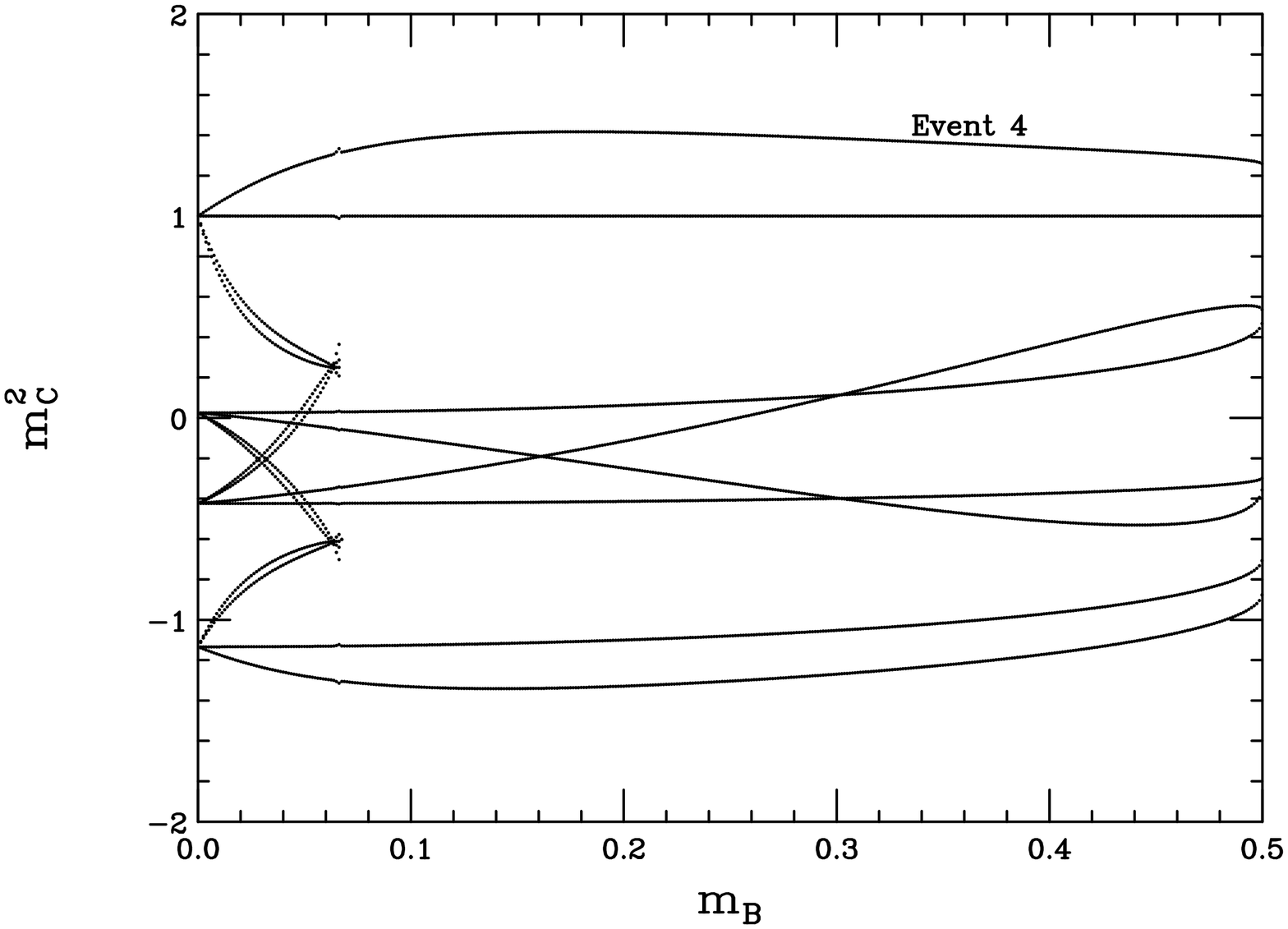}
\caption{Solutions for the mass-squared of $C$ in the double decay chain $C\to B+B',\;B\to A+1,\;B'\to A'+1'$
in 2+1 dimensions, as functions of $m_B$, for four ``typical'' events.   The correct solution is $m_C=1$.   While $m_B$  is varied, each event has fixed decay angles in the parent rest frames. On the left: trajectories of all 16 solutions in the complex mass-squared plane. Each trajectory starts with a cross at $m_B=m_C/2$ (sometimes outside the region shown) and ends with a square at  $m_B=0$.   The intervening points correspond to uniform steps in  $m_B$.  On the right: corresponding plots of the real solutions versus $m_B$. \label{fig:2d}}
\end{center}\end{figure}

In Figure \ref{fig:2d}, we illustrate what happens for four typical events. In the left-hand column, we show all sixteen mass solutions in the complex plane, whereas in the right-hand column we show only the real solutions (the number of solutions is therefore not constant).
In all events, we find that the solutions do indeed coalesce in fours as $m_{B,B^\prime} \rightarrow 0$, one set corresponding in the limit to positive energies for $A$ and $A'$ and the correct mass $m_C$, and the others to unphysical energies for one or both of $A$ and $A'$.
%%%%%%%%%%%%%%%%%%%%%%%%%%%%%%%%%%%%%%
%%%%%%%%%%%%%%%%%%%%%%%%%%%%%%%%%%%%%%

A more realistic example for collider physics was studied in \cite{Cheng:2008mg,Cheng:2009fw}, where pair decays with three visible particles in each chain were considered: $D\to C+3, C\to B+2, B\to A+1$ and similarly for $D'\ldots1'$.
In a single event, there are eight energy-momentum unknowns, together with eight mass-shell constraints and two measured missing transverse momenta, implying two relations between the eight masses along the two chains. If one makes the further hypothesis that the chains are identical, then from two events one obtains four relations between four masses, meaning that one can solve for all masses in the chain.  In \cite{Cheng:2008mg}, it was shown that the system of constraints could be reduced to a single polynomial equation of degree eight in one of the masses. The strategy for dealing with wrong solutions and wrong combinatorics was simply to accept all real solutions and a correlation between right and wrong solutions of the type we describe was observed in numerical simulations. 

Again, it is a simple matter to show that, in the limit that $m_{C,C^\prime} \rightarrow 0$, this eighth order equation reduces to a linear equation. (The same is true if $m_{B,B^\prime} \rightarrow 0$.) Indeed, in each event and in each chain the visible particle 1 is forced to be collinear with the invisible particle $A$, such that we have four equations of the form $p_A = \alpha p_1$, with $\alpha$ unknown. Plugging these into the four transverse missing momentum constraints (two components for each of two events) constitute four equations in the four unknowns $\alpha$, with a unique solution. 

In \cite{Cheng:2008mg}, the relevant masses were taken from the SUSY benchmark point SPS1a, and were $m_{A,B,C} = 97,143,180$ GeV and $m_D=$ either 565 or 571 GeV (for up or down squarks, respectively). Since $m_C$ is substantially less that $m_D$, we therefore conjecture that all eight complex solutions lie close to the the right solution, leading to a correlation between right and wrong solutions over many events. This was indeed observed for the real solutions in \cite{Cheng:2008mg}; the complex solutions were not retained.

Our arguments also permit us to make a useful statement with regard to combinatorics. We have argued that permutations of visible particles 1 and 2 should be irrelevant in the limit that $m_{C,C^\prime} \rightarrow 0$. Now, in the decay chains considered in \cite{Cheng:2008mg}, particles 1 and 2 are either electrons or muons, leading to an eight- (for $2\mu 2e$) or sixteen-fold ambiguity (for $4\mu$ or $ 4e$) per event, or a 64-, 128-, or 256- fold ambiguity per pair of events. But in the limit that $m_{C,C^\prime} \rightarrow 0$, we argue that permutation of  visibles 1 and 2 is irrelevant, in the sense that the solutions obtained after the permutation will be the same as those obtained beforehand. This translates to sixteen irrelevant permutations for a pair of chains and for a pair of events. If we made the na\"{\i}ve assumption that a relevant permutation will lead to a polynomial that has no real solutions, then we would conjecture that one should find precisely sixteen times as many real solutions when one includes combinatorics as compared to when combinatorial ambiguities are removed. In \cite{Cheng:2008mg}, a sample of one hundred events was considered, corresponding to 4,950 event pairs, with 11,662 real solutions in total, without combinatorics. This corresponds to 4,069 event pairs with the minimum number of two real solutions, and 881 with the maximum number of four real solutions. Now, with combinatorics, one must solve 120 times as many degree eight polynomials, but we predict that the number of real solutions will increase by a factor of only sixteen and furthermore that these will be correlated with the right solution. In fact, 185,867 real solutions are obtained in \cite{Cheng:2008mg}, a factor of 15.93 increase compared to the situation without combinatorial ambiguities! Moreover, the pattern of correlation between right and wrong solutions is not changed once one includes combinatorics, as we expect. The mere fact that an odd number of real solutions was obtained in \cite{Cheng:2008mg} once wrong combinations were included shows that one cannot expect perfect agreement: the algorithm used to solve the eighth-order polynomials will, presumably, sometimes fail to converge. Moreover, we only expect the permutations to be truly irrelevant in the limit that $m_{C,C^\prime} \rightarrow 0$; for non-vanishing $m_{C,C^\prime}$ the number of solutions ought to change.  Finally, polynomials obtained from relevant permutations may still have real solutions. Indeed, though they are just random polynomials, they have real coefficients and their zeroes are more likely to lie on the real line than, say, on any other straight line drawn through the origin in the complex plane. 

%%%%%%%%%%%%%%%%%%%%%%%%%%%%%%%%%%%%%%
\subsubsection{Di-leptonic top decays}
%%%%%%%%%%%%%%%%%%%%%%%%%%%%%%%%%%%%%%
Another relevant example of pair cascade decays occurs already in the Standard Model, namely decays of pair produced top quarks in the di-leptonic channel. There, each top quark decays to a bottom quark and a $W$-boson, which subsequently decays to a charged lepton and an invisible neutrino. Since the masses of all particle involved (including the top quark) are relatively well-known, one can attempt to reconstruct the neutrinos' momenta event-by-event. Indeed, there are eight unknowns (the two four-momenta of the neutrinos), together with eight constraints (the six mass-shell constraints and the two missing transverse momentum constraints). Such a reconstruction, if it can be achieved in practice, would be useful, for example, for a likelihood based test of spin correlations\cite{Melnikov:2011ai}. It has previously been shown that the system of constraints can be reduced to a single, quartic equation in one unknown \cite{Sonnenschein:2006ud}. Here we remark only that, in the limit that the $W$-boson mass can be neglected compared to the top quark mass, the system of constraints reduces to a linear equation in a single unknown. (The arguments are much the same as those given above; we do not repeat them here.) Thus, we again expect a correlation between the right and wrong solutions of the quartic, given the fairly small mass ratio between the $W$ and the top. This effect should enhance our ability to measure spin correlations between pairs of top quarks.
%%%%%%%%%%%%%%%%%%%%%%%%%%%%%%%%%%%%%%
\subsection{Massless particle decays}\label{sec:massless}
%%%%%%%%%%%%%%%%%%%%%%%%%%%%%%%%%%%%%%
In \cite{Gripaios:2010hv} search strategies were discussed for composite leptoquarks coupled to third-generation quarks and leptons, which were argued to give a generic and striking signature for models of strongly-coupled electroweak symmetry breaking that can be consistent with constraints from flavour physics \cite{Gripaios:2009dq}. One challenging final state discussed there was the decay of pair produced leptoquarks, each to a top quark and a $\tau$-lepton, with one top decaying hadronically and the other decaying leptonically.

Assuming the leptoquarks are rather massive (existing constraints suggest that their masses should exceed a couple of hundred GeV), then one can neglect the mass of the $\tau$-lepton,
such that the neutrino or neutrinos emitted in the $\tau$ decay may be assumed to be collinear with the visible products of the $\tau$ decay. With this assumption, one is able to solve for the unknown leptoquark mass, given the known masses of the final state particles.
To wit, on the one hand, there are seven unknowns, namely the leptoquark mass, the energy fractions carried off by the neutrinos in the two $\tau$ decays, and the four momentum of the neutrino from the leptonic $W$ decay. On the other hand there are seven constraints, namely the two missing transverse momentum constraints, the mass shell constraints for the two leptoquarks, and the mass shell constraints for the leptonically-decaying top, its daughter $W$, and its daughter neutrino.

It was shown in \cite{Gripaios:2010hv} that this system of seven constraints in seven unknowns can be reduced to a single quartic equation in the energy fraction of one of the neutrinos coming from a $\tau$ decay. It was also observed that there exists a correlation between the right and wrong solutions of the quartic.

We now ask whether this can be understood in the light of the arguments presented here. To do so, let us consider what happens to the four solutions of the quartic as one of the intermediate masses is taken to zero. To begin with, we consider the limit in which the mass of the $W$ may be neglected compared to the mass of the top quark. Then, one may show that the system of equations collapses to a single, linear equation. Indeed, as $m_W \rightarrow 0$, the four-momentum of the neutrino coming from the $W$-decay must be proportional to the four-momentum of the lepton coming from the $W$-decay; the only unknown is the constant of proportionality, or, equivalently, the energy fraction carried off by the lepton. Then, the mass shell constraint for the top quark, together with the two missing transverse momentum constraints, make up a set of three equations that are linear in three unknowns, namely the energy fractions carried off by the neutrinos in the two $\tau$ decays and the $W$ decay. Thus, there is a unique solution for the energy fractions and indeed the other unknowns.

\begin{figure}[ht!]\begin{center}
\includegraphics[angle=0,width=0.8\linewidth]{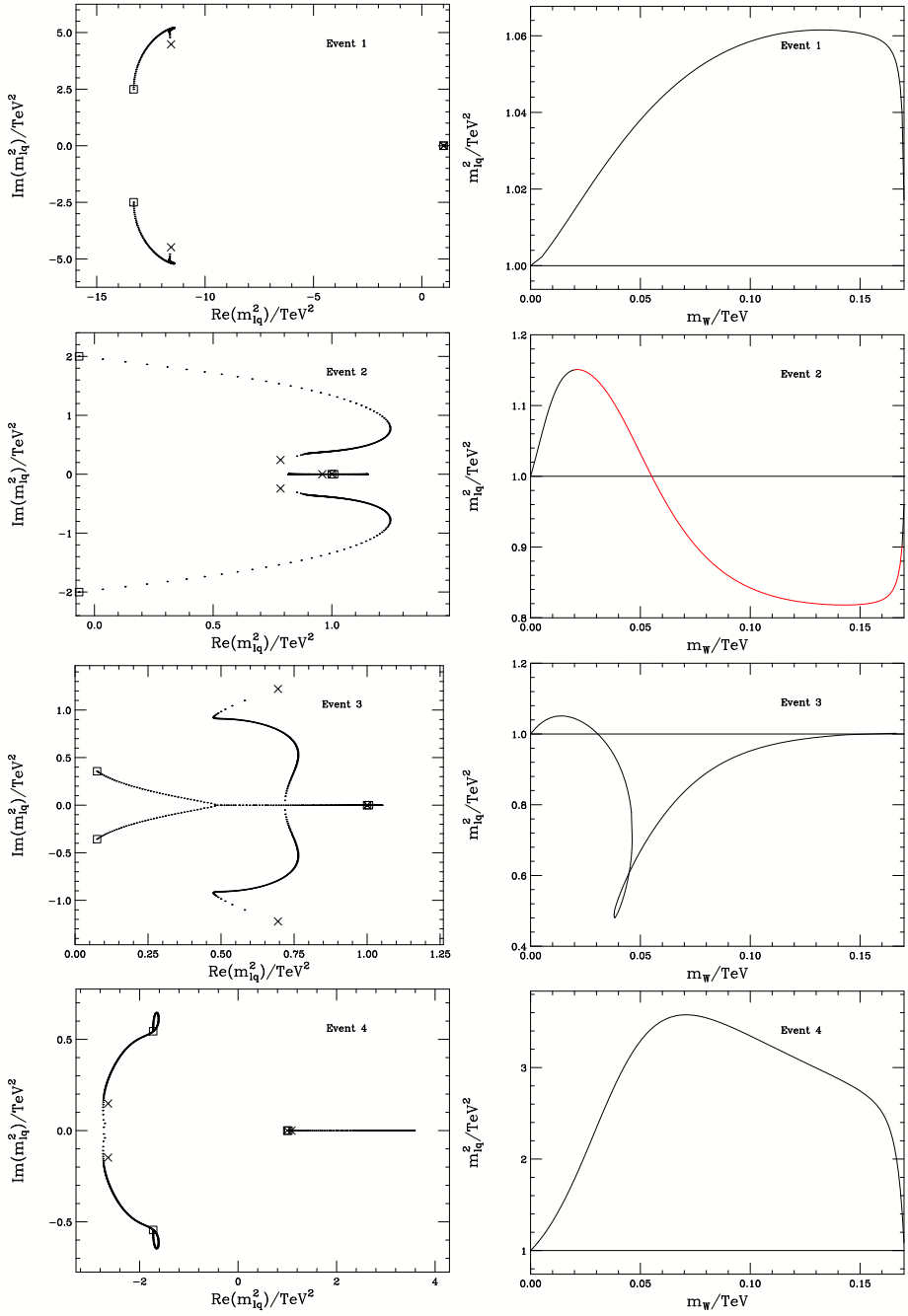}
\caption{Leptoquark mass solutions for four ``typical'' events, as functions of W mass, for a true leptoquark mass of 1 TeV.  On the left: trajectories of all the solutions in the complex mass-squared plane. Each trajectory starts with a cross at $m_W=0.17$ TeV and ends with a square at  $m_W=0$ .   The intervening points correspond to uniform steps in  $m_W$.  On the right: corresponding plots of the real solutions versus $m_W$.  The red portions of the curves correspond to unphysical values of one or both $\tau$ jet energy fractions.\label{fig:lqkW}}
\end{center}\end{figure}

However, it is not the case that the quartic equation collapses to a linear equation. Rather, what happens is that the quartic equation collapses to a cubic equation; one solution of this cubic is, of course, the right solution, whereas the other two solutions are simply not solutions of the the original constraint equations, in the limit. 
This is immediately evident from Figure \ref{fig:lqkW}, where we exhibit numerical solutions of the quartic equation for four typical events. Again, in the left-hand column, we show all four solutions in the complex plane, whereas in the right-hand column we show only the real solutions. In all events, we see that one of the wrong solutions coalesces with the right solution in the limit, but the other two wrong solutions retain a non-vanishing imaginary part in the limit. These complex solutions cannot be solutions of the original system of constraints in the limit, since we saw that the original constraints reduce to three real, linear equations in the three unknown energy fractions, with a unique, real solution. When inserted into the mass shell condition for the leptoquark, these yield a single, real solution for the leptoquark mass.

Hence there is only a two-fold coalescence of solutions in the limit  $m_W \rightarrow 0$, rather than the four-fold coalescence that we might have expected. Nevertheless, this will lead to a correlation between two of the real solutions of the quartic equation.

Figure \ref{fig:lqkW} also shows (in red) that there are solutions which may be discarded on the grounds of being unphysical. In the case at hand, we expect that the energy fractions carried off by invisible particles in decays should not exceed unity. Thus, at least in the limit that measurement errors and combinatorics were under control, one would have grounds for rejecting these solutions, even though they result in real leptoquark masses.
%%%%%%%%%%%%%%%%%%%%%%%%%%%%%%%%%%%%%%
%%%%%%%%%%%%%%%%%%%%%%%%%%%%%%%%%%%%%%
%%%%%%%%%%%%%%%%%%%%%%%%%%%%%%%%%%%%%%
\begin{figure}[ht!]\begin{center}
\includegraphics[angle=0,width=0.8\linewidth]{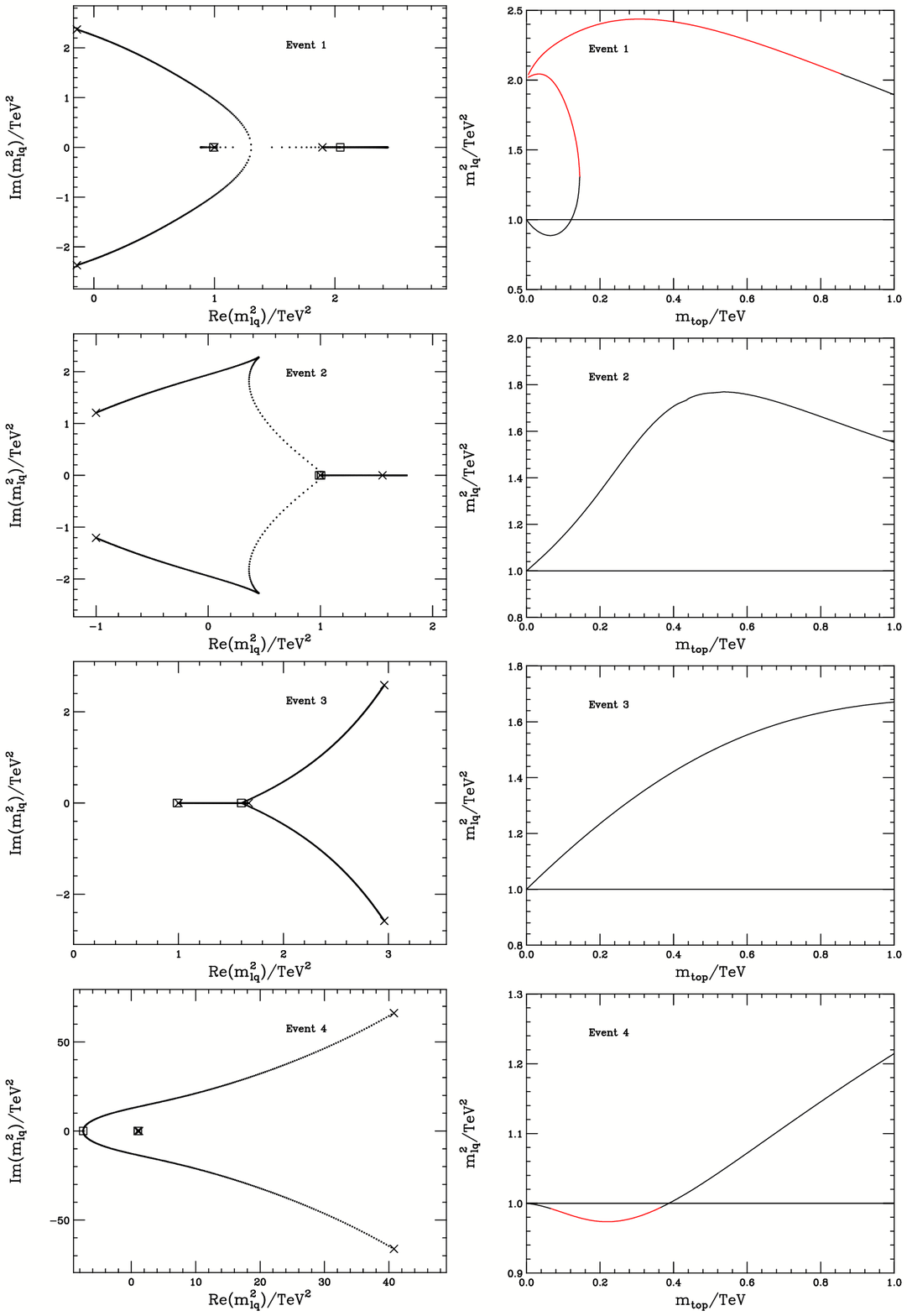}
\caption{Leptoquark mass solutions for four ``typical'' events, as functions of top mass, for a true leptoquark mass of 1 TeV.  On the left: trajectories of all the solutions in the complex mass-squared plane. Each trajectory starts with a cross at $m_{\rm top}=1$ TeV and ends with a square at  $m_{\rm top}=0$ .   The intervening points correspond to uniform steps in  $m_{\rm top}$.  On the right: corresponding plots of the real solutions versus $m_{\rm top}$.  The red portions of the curves correspond to unphysical values of one or both $\tau$ jet energy fractions.\label{fig:lqkT}}
\end{center}\end{figure}
%%%%%%%%%%%%%%%%%%%%%%%%%%%%%%%%%%%%%%%%%%%%%%%%%%%%%%%% 

We may also consider what happens in the limit that the mass of the top quark is assumed to be negligible compared to that of the leptoquark. This obviously implies that the $W$-boson mass may also be neglected, as described above, but it qualitatively changes the behaviour of the solutions, as shown in Figure \ref{fig:lqkT}. Indeed, one can show that neglecting the top mass implies, on its own, that the system reduces from a quartic equation to a quadratic equation. As we have repeatedly described, this implies either that roots go to infinity (which we do not observe), or that roots cease to become solutions of the original system of constraints, or that roots coalesce. For a generic event, we begin with two real and two complex roots. Taking the top mass to zero forces us to have two real roots (since the system reduces to a quadratic equation), but these cannot be the two real roots that we started with, since we know that these coalesce in the limit that the $W$ mass vanishes, which is implied by the vanishing of the top mass.
Thus, the two complex roots must also both become real (and coalesce with each other) in the limit that the top mass vanishes, and indeed this is what we see in all four events in Figure \ref{fig:lqkT}. 

Event 2 in Figure \ref{fig:lqkT} illustrates dramatically the kind of cusp behaviour that we described in Section~\ref{sec:riemann}, arising from branch points of the dual description of the Riemann surface. 

%%%%%%%%%%%%%%%%%%%%%%%%%%%%%%%%%%%%%%
\subsection[Higgs to $\tau\tau$ decay]{\boldmath Higgs to $\tau\tau$ decay}\label{sec:htautau}
%%%%%%%%%%%%%%%%%%%%%%%%%%%%%%%%%%%%%%
In Section~\ref{sec:massless} we considered the decay of a very massive object into a top quark and a $\tau$-lepton, the latter being so highly boosted that it was a good approximation to neglect its mass and treat its decay products as collinear.  If we make the same approximation for a Higgs boson in the favoured mass range $115<m_h<150$ GeV decaying into $\tau\tau$, the kinematics can be reconstructed unambiguously from the visible decay products and the missing transverse momentum.  On the other hand the boost is not so large and, especially after taking into account detector resolution and acceptance, the reconstruction of the Higgs mass may not be optimal.

One can avoid the collinearity assumption by making use of information on the $\tau$ decay vertices.
The most useful and best measured attributes of these are their {\em impact parameters}.  The impact parameter $\vb$  is the displacement of a decay vertex in a direction perpendicular to that of the visible decay momentum, in this case the $\tau$ jet momentum $\vp_j$.  Then the invisible momentum $\vp_\nu$ must lie in the $(\vb,\vp_j)$ plane, so we can write $\vp_\nu=x\vb+y\vp_j$.   For hadronic $\tau$ decays, the invisible momenta are carried by single neutrinos and so their four-momenta are fixed by $x$ and $y$ for each decay.  These four quantities are subject to two linear missing-$p_T$ constraints and two quadratic $\tau$ mass-shell constraints, giving four solutions and hence a fourfold ambiguity in the reconstructed Higgs mass.  However, from our previous arguments the mass hierarchy $m_h\gg m_\tau$ implies that the solutions should tend to be clustered together.

We have investigated this reconstruction method using a sample of 50,000 simulated LHC ($pp$ at 14 TeV) events in which a Higgs boson of mass 130 GeV is produced by vector boson fusion and decays into $\tau\tau$.   The event generator was Herwig++ version 2.5.0~\cite{Bahr:2008pv,Gieseke:2011na}, with parton showering, multiple parton interactions, hadronization and the built-in $\tau$-decay package~\cite{Grellscheid:2007tt}.  The detector simulation was Delphes version 1.9~\cite{Ovyn:2009tx} with its $\tau$-identification algorithm and the ATLAS simulation card.  Vertex information is not provided by Delphes, so we used the hadron-level positions from Herwig++ after gaussian smearing with the r.m.s. values expected for the ATLAS experiment~\cite{Aad:2009wy} (10.5 $\mu$m for the impact parameter).  For the analysis, we demanded two $\tau$-tagged hadronic jets with $p_T>10$ GeV and $|\eta|<2$, resulting in 1467 events remaining after cuts.  

%%%%%%%%%%%%%%%%%%%%%%%%%%%%%%%%%%%%%%
\begin{figure}[htb!]\begin{center}
\includegraphics[width=0.7\linewidth,angle=270]{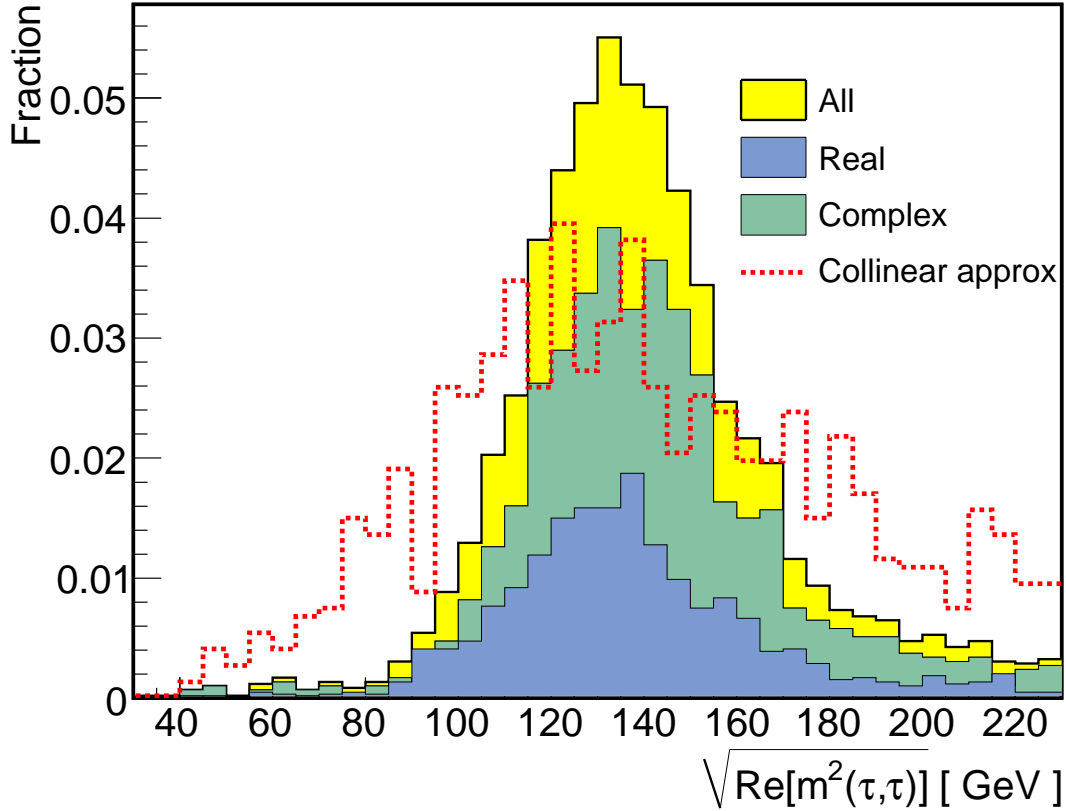}
\caption{Higgs mass reconstructed from simulated detector-level $h\to \tau\tau$ events using impact parameter information, compared with the collinear approximation.\label{fig:htautau}}
\end{center}\end{figure}
%%%%%%%%%%%%%%%%%%%%%%%%%%%%%%%%%%%%%%

Figure~\ref{fig:htautau} shows the Higgs mass reconstructed from the detector-level data using the above method.  All solutions with positive real parts of both reconstructed neutrino energies are included.  We see that after resolution smearing most of the solutions are complex, but the mass resolution from taking their real parts is just as good as that of the real solutions, and substantially better than that of the collinear approximation.  Furthermore, because each solution represents a full reconstruction of the kinematics, there may be scope for further improvement by weighting solutions according to the relevant decay matrix elements.

%%%%%%%%%%%%%%%%%%%%%%%%%%%%%%%%%%%%%%%%%%%%%%%%%%%%%%%%%%%%%%
\section{Discussion and Conclusions}
%%%%%%%%%%%%%%%%%%%%%%%%%%%%%%%%%%%%%%%%%%%%%%%%%%%%%%%%%%%%%%
Reconstruction of missing energy events may be important for many physics
analyses at colliders. Even in the Standard Model, missing energy is ubiquitous, in the form of neutrinos, which are invisible in the detectors.
Reconstruction of Standard Model events may be useful for, for example, improved measurements of the top quark mass, or for identifying the presence of spin correlations in pair production of top quarks.

Reconstruction may prove to be even more important for physics beyond the Standard Model, not only because we hope to see missing energy in events
involving dark matter particles, but also because it is not so easy in the case of new physics to specify the signal hypothesis, that is to say, the lagrangian.

A significant complication affecting reconstruction of energies, momenta, and masses in missing energy events at colliders is the presence of multiple solutions. As we have seen, the number of solutions can be large (sixteen in one of the examples we considered). This is compounded by the presence of combinatorial ambiguities and measurement errors, which further increase the number of solutions and make it less easy to decide which of the multiple solutions is the correct one.

In the worst case scenario, one would have to accept all solutions, correct or incorrect, real or complex, with the risk that the ``signal'' of correct solutions would be overwhelmed by the ``background'' of incorrect solutions.
Here, we have shown that this problem is mitigated by the existence of mechanisms by which the incorrect solutions are correlated with the correct ones. Specifically, we found that correct and incorrect solutions may coincide in the limit that intermediate masses in cascade decay chains either are negligible, or are degenerate with the masses of particles further up the chain, such that the emitted particles are either collinear, or soft, respectively. Furthermore, these same limits can also lead to combinatorial ambiguities becoming irrelevant, in the sense that the same solutions are obtained before and after a permutation of particles. The correlations between correct and incorrect solutions, which are perfect at either end of the interval of possible intermediate particle masses, persist throughout the intermediate mass interval.

We saw that these phenomena have a natural description in terms of the theory of Riemann surfaces, and studied several examples relevant to colliders, for processes both in and beyond the Standard Model.
We hope that our results provide some insight into the general problem of reconstructing events with missing energy and that they will be useful to those who seek to do so in today's colliders.

More specifically, we have shown that the closeness of correct and incorrect solutions means that complex solutions can occur even in the presence of small measurement errors. Whilst existing analyses discard complex solutions (on the grounds that they must correspond to events with large mismeasurements) we recommend that future analyses retain all solutions, with a consequent increase in the available statistics. As discussed in the introduction and Section~\ref{sec:htautau}, the ongoing searches at the LHC for Higgs bosons in decays to pairs of $\tau$ leptons would seem to be a good place to begin.
%%%%%%%%%%%%%%%%%%%%%%%%%%%%%%%%%%%%%% 

\section*{Acknowledgments}
BMG thanks N. Orantin for discussions.  Part of the work of BMG and BW was performed at the Kavli Institute for Theoretical Physics, University of California, Santa Barbara, supported in part by the National Science Foundation under Grant No. NSF PHY05-51164.  BW also acknowledges the support of a Leverhulme Trust Emeritus Fellowship, and thanks the CERN Theory Group for hospitality.

%%%%%%%%%%%%%%%%%%%%%%%%%%%%%%%%%%%%%%%
\providecommand{\href}[2]{#2}\begingroup\raggedright\endgroup

%%%%%%%%%%%%%%%%%%%%%%%%%%%%%%%%%%%%%%
\end{document}